\documentclass[aps,prd,amsmath,amssymb,10pt,letterpaper,balancelastpage,showpacs,superscriptaddress,nofootinbib,floatfix,notitlepage,longbibliography,twocolumn
]{revtex4-2}
\usepackage[utf8]{inputenc}
\usepackage{textgreek}
\usepackage[normalem]{ulem}
\usepackage{comment}
\usepackage{graphics}
\usepackage{bm}
\usepackage{graphicx}
\usepackage{color}
\usepackage{amsmath}
\usepackage{amssymb}
\usepackage{mathrsfs}
\usepackage{graphicx}
\usepackage{caption}
\usepackage{subcaption}
\usepackage{accents}
\usepackage{adjustbox}
\usepackage{multirow}
\usepackage{diagbox}
\usepackage{stackengine}
\usepackage{float}
\usepackage{ulem}

\usepackage[dvipsnames]{xcolor}
\definecolor{red}{rgb}{0.9, 0,0}
\definecolor{cerulean}{rgb}{0., 0.42,0.9}
\definecolor{navy}{rgb}{0.05, 0.05,0.8}

\usepackage[colorlinks]{hyperref}
\hypersetup{
  colorlinks = true,
  citecolor = red,
	linkcolor = navy
}
\usepackage{slashed}

\newcommand{\Eq}[1]{Eq.~(#1)}

\newcommand{\rhoDM}{\rho_{\text{DM}}}

\newcommand{\vbold}{\mathbf{v}}

\newcommand{\vmax}{v_{\text{max}}}

\newcommand{\vmin}{v_{\text{min}}}
\newcommand{\sxn}{\sigma_{\chi n}}

\newcommand{\mx}{m_{\chi}}

\newcommand{\kms}{\text{km}/\text{s}}

\newcommand{\gpg}{\text{g}/\text{g}}
\newcommand{\Nsignal}{N_{\rm signal}}
\newcommand{\Eobs}{E_{\rm obs}}

\begin{document}
\title{Testing the inelastic dark matter interpretation of the LZ event with heavy-element experiments}

\author{Peter W.~Graham}
\affiliation{Leinweber Institute for Theoretical Physics, Department of Physics, Stanford University, Stanford, CA 94305, USA}
\affiliation{Kavli Institute for Particle Astrophysics and Cosmology, Department of Physics, Stanford University, Stanford, CA 94305, USA}

\author{Juan P. \surname{Luengas}}
\email{jluengas@udel.edu}
\affiliation{Department of Physics and Astronomy, University of Delaware, Newark, DE 19716, USA}

\author{Harikrishnan Ramani}
\affiliation{Department of Physics and Astronomy, University of Delaware and the Bartol Research Institute, Newark, DE 19716, USA}

\author{Samuel S.~Y.~Wong}
\email{samswong@uw.edu}
\affiliation{Department of Physics, University of Washington, Seattle, WA 98195, USA}

\begin{abstract}
We consider the inelastic dark matter interpretation of the recent 248~keV event observed by LZ and identify different experiments that could test it. Details of the underlying model and the ability of complementary experiments to test it depend significantly on the high-velocity tail of the dark matter distribution. We consider several example velocity distributions motivated by the impact of the Large Magellanic Cloud (LMC) in order to map out the effects of this dependence. For example, we illustrate the type of high-velocity tail that would make the thermal Higgsino viable despite stringent constraints from solar neutrinos: Reducing the flux of the tail relative to simulation results improves the Higgsino fit, while extending the tail to higher velocities at fluxes below current simulation resolution evades the solar neutrino bound. This motivates more detailed simulations at higher resolution to understand the tail better. Even if xenon experiments confirm the signal event, complementary experiments are still required to test whether the underlying model is inelastic dark matter and to break degeneracies in the measurement of dark matter parameters such as mass and mass splitting. Heavy-element paleodetectors could cover most of the parameter space relevant to the LZ event.  Their use of different elements could determine the dark matter mass and splitting, while their unique sensitivity to the history of the dark matter distribution could test the signal's LMC origin. Luminous detection in large neutrino detectors with heavy-element supplements could directly test the inelastic nature of the signal and measure the mass splitting, while its sensitivity to the radiative decay rate could provide strong evidence for the Higgsino interpretation.
\end{abstract}

\maketitle

\tableofcontents

\section{Introduction}

The LZ experiment recently observed a single nuclear-recoil event at $E_R = 248~\mathrm{keV}$~\cite{LZ:2026axp}. This exciting development has raised hope for finally understanding the particle nature of dark matter (DM)~\cite{Yamashita:2026ump,McCabe:2026crm,Jeesun:2026vzo,Unwin:2026rdp,deLima:2026shq,Gu:2026vto,Baer:2026fpy,Lee:2026wof,Liang:2026coz,Das:2026uyy,Alhazmi:2026efz,Okada:2026eol,Du:2026lpa,Kannike:2026qyl,Yuan:2026djt,Zhu:2026dag,Aghaie:2026vsu,Asadi:2026iot,Khan:2026nwp,He:2026hqz,Fan:2026hzw,Chattaraj:2026fxn,Kumar:2026lgi,Heikinheimo:2026kwp,Lee:2026zbr,Mahapatra:2026glu,Barman:2026omh,Das:2026buc,Ahmed:2026kan,He:2026idw,Lian:2026hpm,Baer:2026yrt,Uttayarat:2026isp,Arcadi:2026kev,An:2026pkc,Xing:2026civ,Okada:2026fef,Ahmed:2026com,De:2026win,Sheng:2026tqt,Chauhan:2026udz,Sannino:2026hkc,Khan:2026osp,Jung:2026otm,Gemmell:2026yaw,Tang:2026lgu,Su:2026rwz,Palmisano:2026kuj,Ge:2026xax,Lee:2026xxh,Le-Yaouanc:2026djt,Yang:2026wpb,Brdar:2026ukx,Kalita:2026hvb,Qi:2026vyp}. The observed recoil energy is much higher than the conventional weakly interacting massive particle (WIMP) recoil window. Furthermore, there are no corresponding low-energy events. Such a spectrum is unlikely to occur for ordinary elastic spin-independent scattering, but arises naturally for
inelastic dark matter, whose scattering against nucleons requires a transition
between two nearly degenerate states separated by a mass splitting $\delta$~\cite{Tucker-Smith:2001myb}. The kinematic threshold imposed by the inelasticity suppresses low-energy recoils. 

Of particular interest is the Higgsino in the minimal supersymmetric standard model (MSSM)~\cite{EnhancingHiggsinoDM_2024,Nagata:2014wma}. In the heavy-gaugino
limit, the two neutral Higgsino states form a pseudo-Dirac fermion, which has an off-diagonal Z coupling. The nuclear scattering cross section is then fixed to the electroweak scale. Furthermore, thermal freeze-out fixes the Higgsino mass to $\mx=1.1$~TeV, modulo possible non-standard cosmologies. The Higgsino interpretation of the LZ event~\cite{Fan:2026kxx,Freese:2026sga,Wu:2026nhi,
Yin:2026jnn,Du:2026guj,Cheung:2026byg,Bisal:2026khf,Maier:2026qpr,Langhoff:2026ujr,Unwin:2026qep} is thus highly predictive: for a mass near a TeV, the main freedom is the mass splitting $\delta$.

However, there exist strong constraints on this interpretation. Solar capture and annihilation exclude $\delta \lesssim 566~\mathrm{keV}$ for a thermal Higgsino~\cite{Pospelov:2026ewn}, while the lack of additional LZ events at energy above 248~keV further restricts the allowed parameter space~\cite{Rodd:2026tyn}. On the other hand, inelastic dark matter is extremely sensitive to the high-velocity tail of the local DM distribution, where the Standard Halo Model (SHM) is least trustworthy~\cite{Lisanti_TASI, SMH_Deviation_4}. In particular, the recent pericenter approach of the Large Magellanic Cloud (LMC) roughly $50~\mathrm{Myr}$ ago can enhance this tail~\cite{Besla:2019xbx, Smith-Orlik:2023kyl, Donaldson_2022}, making its effect particularly relevant for direct detection of inelastic DM, as first pointed out by some of us in Refs.~\cite{EnhancingHiggsinoDM_2024,Graham:2026ivn}.

Present simulations of the LMC's effect may not fully resolve its lowest-flux, highest-velocity
component~\cite{Smith-Orlik:2023kyl}. We therefore ask whether possible high-velocity tails of the
distribution can reconcile the LZ event with both the solar-neutrino bound and the absence of additional high-energy recoils. Furthermore, this event may originate from a different inelastic dark matter model rather than the Higgsino. Even in this case it is important to ask how small changes in the high-velocity tail affect the best-fit parameters of that model. We find that a high-velocity tail with \emph{less} flux than some of the results from simulations and/or an extended maximum velocity with a tiny flux below the simulation resolution gives a better fit to the thermal Higgsino. This finding motivates more detailed simulations at higher resolution.

Even if LZ or another xenon experiment were to confirm the signal event, there is still too much degeneracy to establish the nature of dark matter, as different DM models can produce similar spectra. Therefore, we need complementary experiments with different target nuclei and qualitatively different observables to test the inelastic DM hypothesis.

A natural direction is to use heavy nuclei. For TeV-scale dark matter, nuclei heavier than xenon have larger center-of-mass kinetic energy and can therefore access larger velocity phase space and larger mass splittings. Heavy-element experiments are thus powerful probes of generic inelastic DM, beyond just the Higgsino interpretation. 

One possibility is the heavy-element paleodetectors~\cite{Graham:2026ivn}, which
search for persistent damage tracks in ancient minerals accumulated over geological timescales. We find that minerals containing Pb can probe much of the parameter space relevant to the LZ event. A major advantage of paleodetectors is their flexibility in the choice of target elements compared to conventional direct detection. We will show that just two different elements can in principle determine $\mx$ and $\delta$. Paleodetectors are also uniquely sensitive to the history of the DM velocity distribution. A series of samples of different ages can therefore determine whether a signal is produced by the enhanced high-velocity tail due to the recent LMC encounter; this information is inaccessible to any real-time experiment.

A second possibility is luminous dark matter detection~\cite{Feldstein:2010su,Eby:2019mgs}. If the excited dark matter
state has a radiative decay,
$\chi^\prime \rightarrow \chi+\gamma$,
then up-scattering on heavy nuclei in the Earth can be followed by emission of
a monochromatic photon with energy $E_\gamma = \delta$. For a macroscopic
decay length larger than the size of the detector and smaller than the size of the
Earth, this photon can be observed in a large-volume underground detector such
as JUNO~\cite{JUNO_Conceptual}. This mechanism applies to a broad class of inelastic DM models with a visible excited-state decay. The Higgsino provides a particularly
predictive realization because its radiative decay rate is calculable.

The luminous rate can be further enhanced by placing dedicated volumes of
lead (or other heavy elements such as depleted uranium) around the detector~\cite{EnhancingHiggsinoDM_2024}. Such a supplement not only increases the
up-scattering rate but also introduces sensitivity to the decay length itself, allowing for a direct confirmation of whether an observed signal is caused by the thermal Higgsino.

The rest of this paper is organized as follows. In Section~\ref{sec:LZ}, we explore the compatibility of the LZ event with various modifications of the velocity distribution. In Section~\ref{sec:experiments}, we review the heavy-element experiments and show their reach into the parameter space compatible with the LZ event, as well as discuss their unique features that allow for testing the inelastic DM hypothesis. We conclude in Section~\ref{sec: conclusion}.

\section{Inelastic DM interpretation of the LZ event}
\label{sec:LZ}

\subsection{Signal rate}

The differential DM-nucleus scattering
rate per target mass with respect to recoil energy is given by
\begin{equation} \label{eq:dRdER}
\frac{dR}{dE_R}
=
\frac{\rhoDM \sxn}
{2 m_\chi \mu_n^2} \sum_A w_A Q_A^2
F_A^2(E_R) \,
\eta\!\left(v_{\min}(E_R)\right) ~,
\end{equation}
where $\sxn$ is the DM-nucleon cross section, $\rhoDM = 0.3~{\rm GeV/cm^3}$ is the DM density, $\mu_n$ is the DM-nucleon reduced mass, $F_A(E_R)$ is the Helm form factor~\cite{EnhancingHiggsinoDM_2024}, and we sum over the relevant target isotopes with mass number $A$ and mass fraction $w_A$. We use a DM-nucleon coupling consistent with the Z-mediated case,
\begin{align}
    Q_A^2 = \left[ N - \left(1-4\sin^2\theta_W\right) Z\right]^2 ~,
\end{align}
where $N=A-Z$ is the neutron number, $\theta_W$ is the Weinberg angle, and the accidentally small factor $1-4\sin^2\theta_W \approx 0.11$ suppresses the DM-proton coupling.\footnote{This is in contrast to the equal-coupling assumption $Q_A^2=A^2$ sometimes used in the inelastic dark matter literature, e.g., Ref.~\cite{EnhancingHiggsinoDM_2024}.} For the Higgsino, the Z-mediated cross section is given by\footnote{An incorrect extra factor of $1/4$ has been propagated in the literature~\cite{Pospelov:2013nea}.}
\begin{align}
    \sxn = \frac{G_F^2 m_n^2}{2\pi} \approx 7 \times 10^{-39}~\text{cm}^2 ~.
\end{align}
The dependence on the DM velocity distribution is contained in the
time-averaged lab-frame halo integral~\cite{EnhancingHiggsinoDM_2024},
\begin{equation}
\eta(v_{\min})
=
\int_{v \geq v_{\min}}
d^3v\,
\frac{f^{\rm det}(\mathbf{v})}{v} ~,
\end{equation}
where the minimum DM-velocity required for a given recoil energy $E_R$ is
\begin{equation} \label{eq:vmin}
    v_{\min}(E_R)=\frac{1}{\sqrt{2m_A E_R}}\left(\frac{m_A}{\mu_A}E_R+\delta\right) ~,
\end{equation}
where $\mu_A$ is the DM-nucleus reduced mass.

\begin{figure}[t!]
    \centering
    \includegraphics[width=0.5\textwidth]{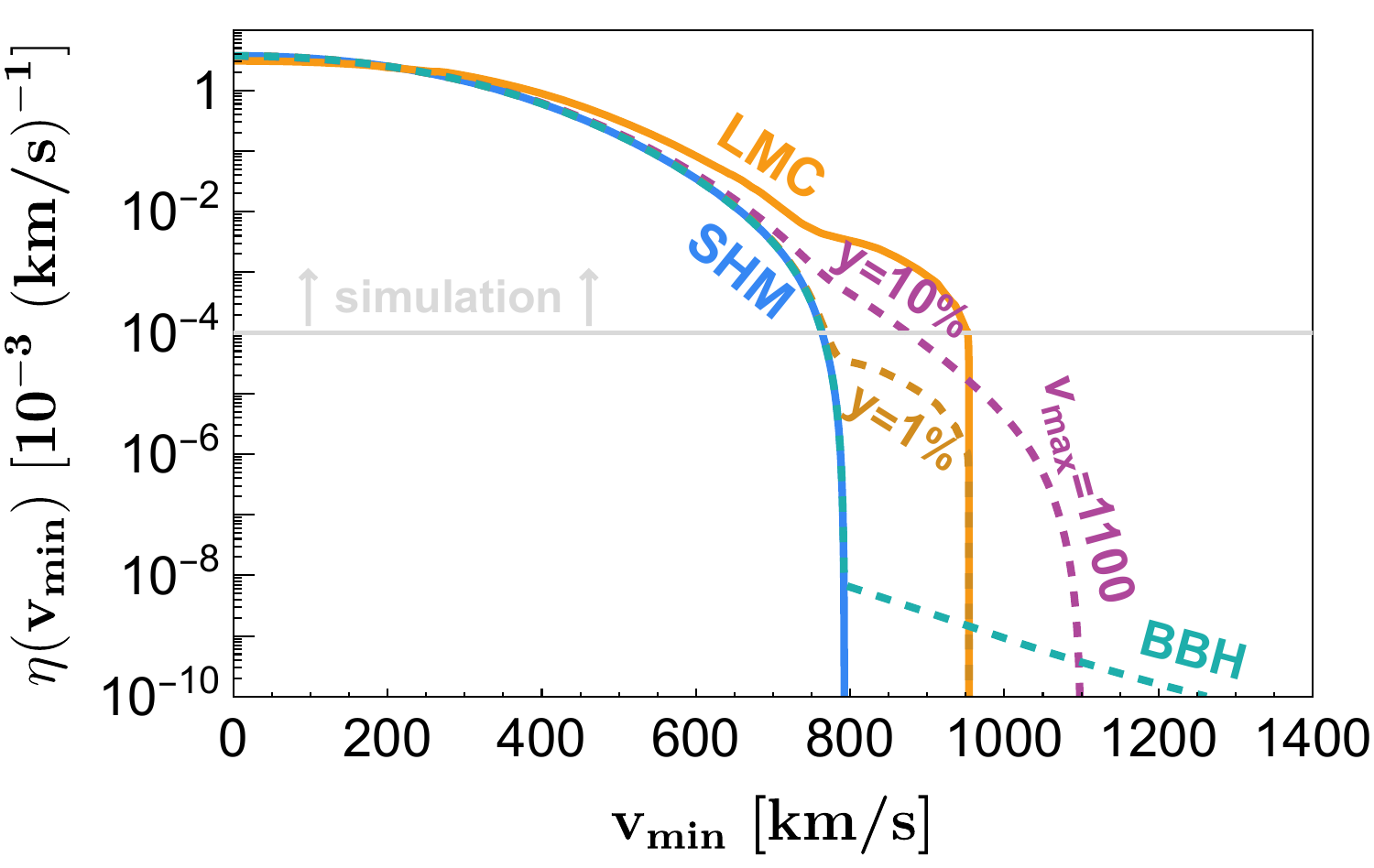}
    \caption{Halo integrals for the Standard Halo Model (blue); the LMC model (orange) from Auriga simulation~\cite{Smith-Orlik:2023kyl}; the LMC model but whose tail has a flux at the percent-level (gold dashed); the LMC model with a 10\%-flux tail and extended to $\vmax=1100$~km/s (purple dashed); a tail due to DM boosted by binary black holes in the galaxy (teal dashed), assuming $\gamma = 1.5$ in the NFW model~\cite{Acevedo:2026xol}. The gray line indicates the smallest flux reported by simulations~\cite{Smith-Orlik:2023kyl}.}
    \label{fig:halo}
\end{figure}

\subsection{High-velocity tail from the LMC}

A common assumption for the DM velocity distribution is the Standard Halo Model (SHM)~\cite{drukier:1986tm}, a galactic-frame Maxwell-Boltzmann distribution truncated at the Milky Way (MW) escape velocity,
\begin{align}
    f^{\rm gal}_{\rm SHM}(v) \propto e^{-v^2/v_0^2} \Theta(v_{\rm esc}-v) ~, 
\end{align}
where $v_0 = 220$~km/s and $v_{\rm esc}=544$~km/s. After boosting to the detector frame and averaging over the Earth's rotation around the Sun (see Appendix~\ref{app:detector_frame}), we denote the distribution by $f^{\rm det}(v;v_0,\vmax)$, where the maximum DM velocity in the detector frame is $\vmax=793~\kms$ for the SHM.

However, recent work~\cite{Besla:2019xbx, Smith-Orlik:2023kyl, Donaldson_2022} in galactic dynamics shows that the high-velocity tail of the distribution may be dominated by the effect of the close pericenter approach of the Large Magellanic Cloud (LMC) $ \sim 50$~Myr ago~\cite{Patel_2020} in the opposite direction to solar motion, possibly resulting in a population of unbound DM. It was first pointed out in Refs.~\cite{EnhancingHiggsinoDM_2024,Graham:2026ivn} by some of us that this high-velocity tail from the LMC could significantly alter direct detection prospects of inelastic DM, and the Higgsino in particular.

Since inelastic scattering has the largest available phase space when the relative velocity between the Earth and the DM is greatest, the fact that the LZ event was observed in June near the annual maximum is suggestive~\cite{LZ:2026axp}. This is true not only in the SHM but also for the LMC model, precisely because the LMC pericenter approach was in the opposite direction to present-day solar motion, and the DM it accelerated or left behind also inherited this direction~\cite{Besla:2019xbx,EnhancingHiggsinoDM_2024,Fan:2026kxx}.

We consider the LMC model (orange curve in Fig.~\ref{fig:halo}) from the Auriga cosmological magneto-hydrodynamical simulations~\cite{Auriga}.\footnote{What we call the ``LMC model" is a simulation that incorporates the effect of both MW and the LMC, and is known as halo 13 in Ref.~\cite{Smith-Orlik:2023kyl}, also known as Auriga halo 25.} Ref.~\cite{Smith-Orlik:2023kyl} only reports a flux down to $\eta = 10^{-7}~(\text{km}/\text{s})^{-1}$ (gray line in Fig.~\ref{fig:halo}), which gives $\vmax = 954~\kms$. We speculate that this maximum velocity could possibly be a consequence of the limit of the simulation (e.g., the simulation running out of DM particles). A more computationally expensive simulation might yield an even faster tail at smaller flux. In general, the shape of the high-velocity tail is uncertain, as there is significant variation among different simulations and there is significant uncertainty in attempts to determine it empirically (see e.g.~\cite{Smith-Orlik:2023kyl, Zhang:2026qnl}).

To investigate what velocity tail is required for the LZ event to be compatible with a thermal Higgsino, we parametrize the unknown extended velocity tail in two ways as defined below. First, we consider only varying the flux of the tail while fixing the maximum velocity in the LMC simulation. To this end, we consider the linear interpolation between the SHM and the LMC model:
\begin{align} \label{eq:interpolation_w}
    f^{\rm det}_y(v) =  (1-y)f^{\rm det}_{\rm SHM}(v) + y \, f^{\rm det}_{\rm LMC}(v) ~,
\end{align}
where $y=1$ corresponds to the full LMC model, including the MW contribution. As we will show, a velocity tail with \emph{less} DM flux than what simulation shows improves the fit to the LZ event. In effect, we consider the $y=1\%$ model (gold dashed curve in Fig.~\ref{fig:halo}), which we will refer to as $\text{LMC}_{y=1\%}$.

Note that our interpolation parameter $y$ differs from the LMC mass fraction $w$ used in Refs.~\cite{Rodd:2026tyn,Langhoff:2026ujr}. The full simulation model (halo 13 in Ref.~\cite{Smith-Orlik:2023kyl}) already has a small LMC mass fraction $0.26\%$, so our prescription corresponds to $w=y\times 0.26\%$. The $y=1\%$ model therefore corresponds to $w=0.0026\%$, which is 100 times smaller than the usual benchmarks in the literature. We will show that even such a tiny flux has a significant impact and actually improves the Higgsino fit to the LZ event.

Second, we consider the effect of varying the maximum velocity as follows. Notice that taking the analytic form of the SHM~\cite{Lee:2013xxa} but using the following parameters roughly reproduces the LMC model, $f^{\rm det}_{\rm LMC}(v) \approx f^{\rm det}(v; v_0=260~\kms,\vmax=954~\kms)$, including its preferred parameter region relevant to the LZ event. Then we can vary the highest velocity tail by simply choosing a different $\vmax$, and vary its flux using the linear interpolation against the SHM as in \Eq{\ref{eq:interpolation_w}}. Motivated by a fit to the LZ event, we choose the representative values $\vmax =1100~\kms$ and $y=10\%$, and we will henceforth refer to this model as $\text{LMC}(1100)_{y=10\%}$ (purple dashed curve in Fig.~\ref{fig:halo}). While this extended velocity is speculative, we note that its flux is an order of magnitude below the smallest flux reported by simulations.

Finally, we briefly consider the possibility of a low-flux, high-velocity tail due to DM boosted by binary black holes (BBH) in the galaxy~\cite{Acevedo:2026xol} (teal dashed curve in Fig.~\ref{fig:halo}).  This serves as an example of what the consequences might be for any rare processes in the galaxy that boost DM up to very high velocities.

\begin{figure}[t!]
    \centering
    \includegraphics[width=0.5\textwidth]{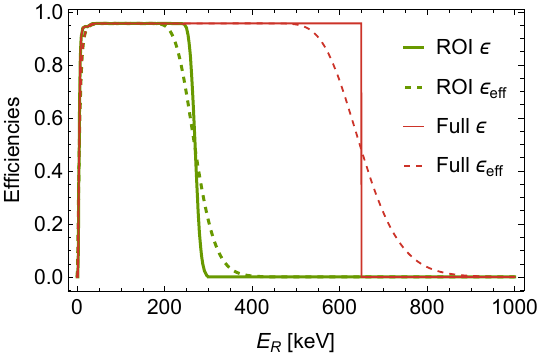}
    \caption{Detection efficiency $\epsilon(E_R)$ and effective efficiency $\epsilon_{\rm eff}(E_R)$, for the region of interest (ROI) vs full energy window.}
    \label{fig:efficiency}
\end{figure}

\subsection{Energy smearing and efficiency window}

We model the total number of signal events by
\begin{align}
  \Nsignal = \mathcal{E} \int_0^{\infty} dE_R \frac{dR}{dE_R} \epsilon_{\rm eff}(E_R) ~,
\end{align}
where $\mathcal{E}=2.84~\text{ton-year}$ is the experimental exposure at LZ~\cite{LZ:2026axp}, and the effective efficiency $\epsilon_{\rm eff}(E_R)$ is the convolution of the detection efficiency $\epsilon(E_R)$ and an energy-smearing Gaussian with an energy-dependent width fitted to the experimental uncertainties~\cite{Rodd:2026tyn}, 
\begin{align}
\sigma_E(E_R) = \sqrt{(23~\text{keV})^2 \frac{E_R}{248~\text{keV}}+(0.093 E_R)^2} ~.
\end{align}
This energy smearing allows for the true recoil energy to be above the efficiency cutoff point, alleviating the tension of the event with larger mass splittings.

Because LZ's calibration events do not include neutron sources that cause recoils in excess of $250$~keV, the experiment conservatively reports a detection efficiency with a region of interest (ROI) roughly around 0--250~keV~\cite{LZ:2026axp}, where a tentative signal event was observed at $E_{\rm obs}=248$~keV. However, this calibration has been validated against simulation up to at least $650$~keV~\cite{LZ:2026axp} and there is no reason to assume the actual detector efficiency cuts off around 250 keV. In most of our plots we will assume that the efficiency remains near 1 all the way up to this energy (the ``full efficiency")\footnote{Note that the full efficiency window we consider differs from that of Ref.~\cite{Rodd:2026tyn}, which stops earlier at $590$~keV and leaves a gap in an intermediate range before $350$~keV, which, in fact, contains valid data~\cite{HallPrivate}.}, although we will also display some plots with the more conservative assumption that we can only use the ROI defined in the LZ paper.  The fact that no additional event was observed in this full energy window up to $650$~keV~\cite{LZ:2026axp} sets an additional constraint on the inelastic DM parameter space compatible with the observed event~\cite{Rodd:2026tyn}. See Fig.~\ref{fig:efficiency} for our two example efficiency curves and the effect of energy smearing.

\begin{figure*}[t!]
    \centering
    
    \begin{subfigure}{0.49\textwidth}
        \centering
        \includegraphics[width=\textwidth]{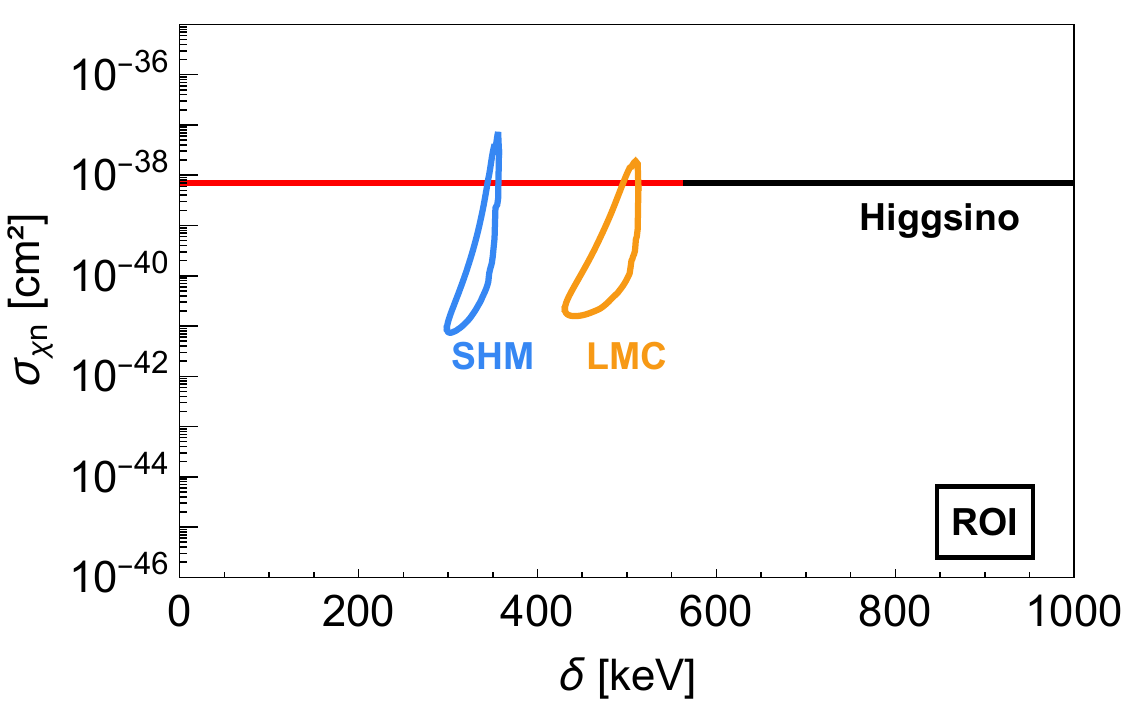}
        \caption{}
        \label{fig:SHM_LMC_ROI}
    \end{subfigure}
    \hfill
    \begin{subfigure}{0.49\textwidth}
        \centering
        \includegraphics[width=\textwidth]{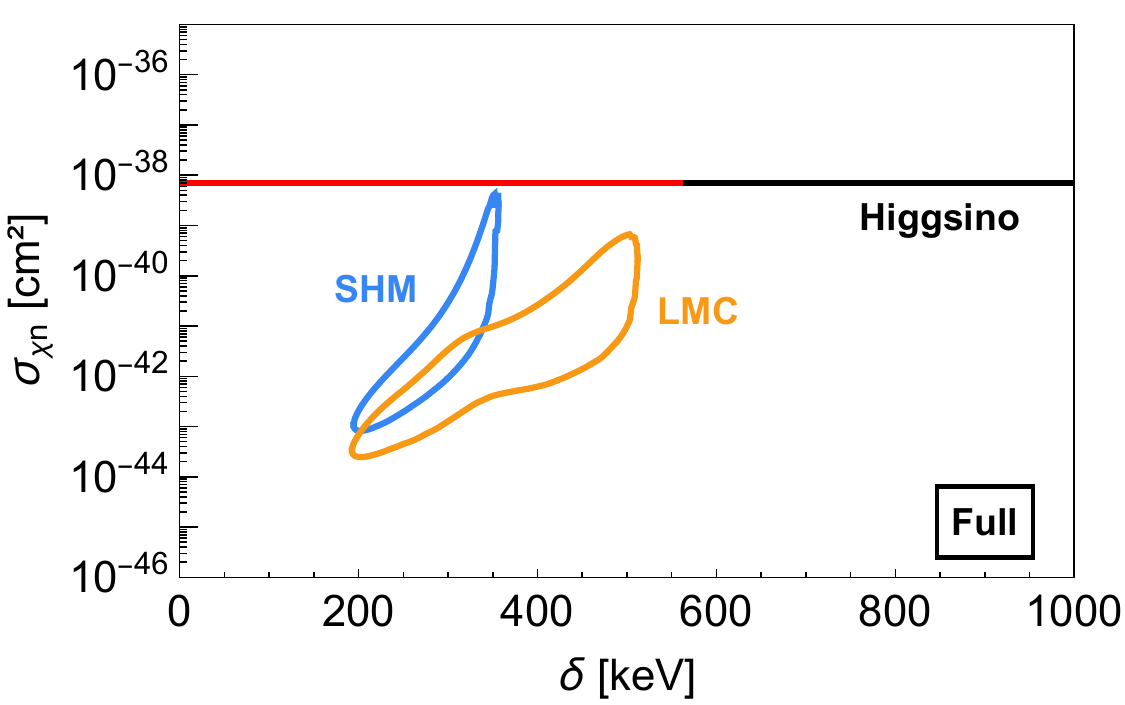}
        \caption{}
        \label{fig:SHM_LMC_full}
    \end{subfigure}
    
\caption{The $90\%$ confidence likelihood contours consistent with the observed event by LZ in the DM-nucleon cross section $\sxn$ vs mass splitting $\delta$ parameter space at fixed DM mass $\mx=1.1$~TeV, for both the Standard Halo Model and the LMC model, using efficiency windows according to (a) LZ's region of interest (ROI) around 0--250~keV; (b) the full window 0--650~keV. The Higgsino cross section is represented by the horizontal black line, where the red portion ($\delta<566$~keV) is ruled out by the solar neutrino constraint~\cite{Pospelov:2026ewn}.}
    \label{fig:contour_SHM_LMC}
\end{figure*}

\begin{figure*}[t!]
    \centering
    
    \begin{subfigure}{0.49\textwidth}
        \centering
        \includegraphics[width=\textwidth]{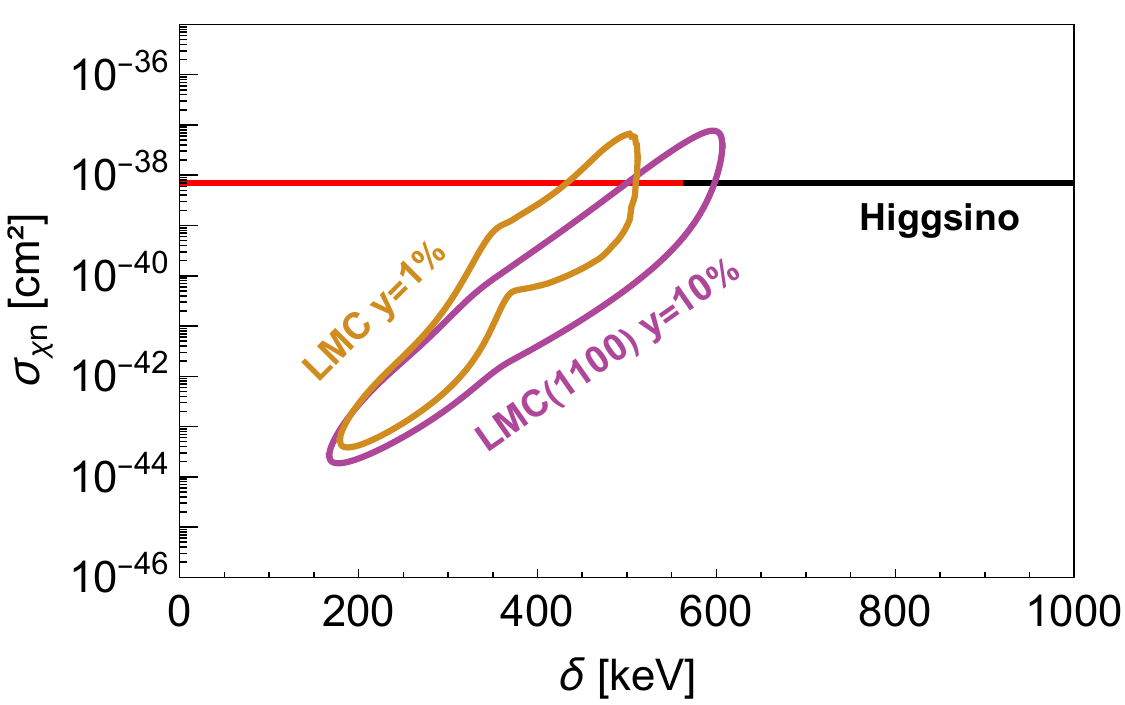}
        \caption{}
        \label{fig:contour_LMC_extended}
    \end{subfigure}
    \hfill
    \begin{subfigure}{0.49\textwidth}
        \centering
        \includegraphics[width=\textwidth]{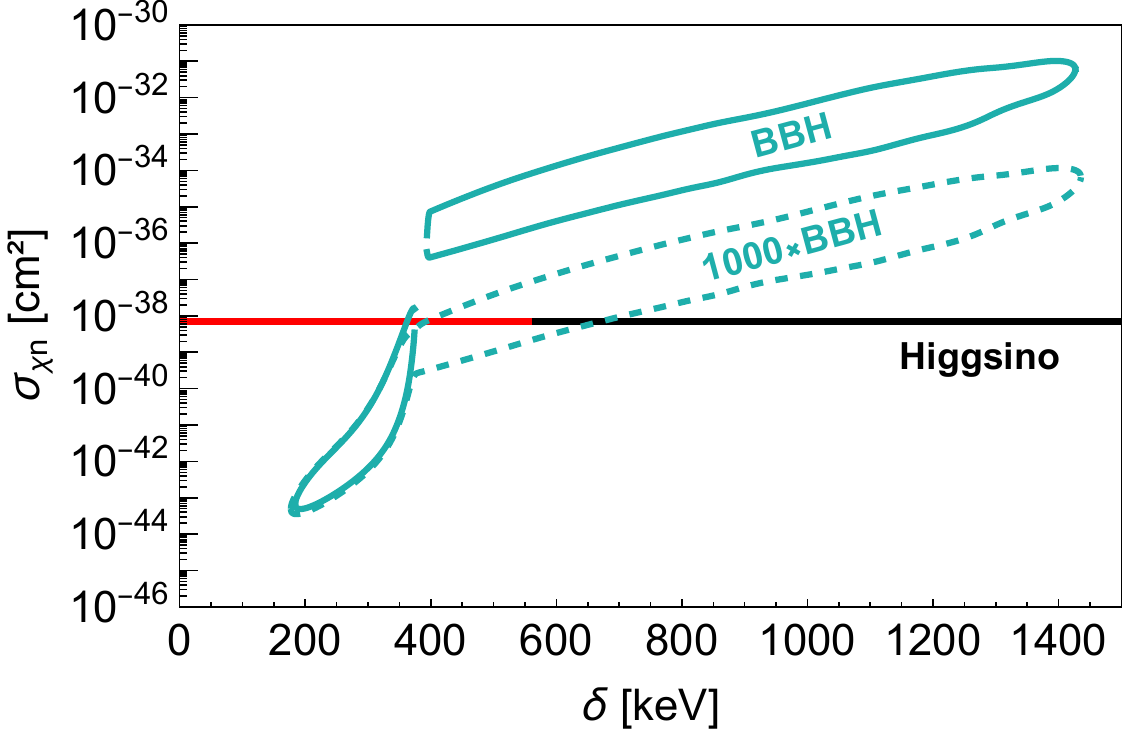}
        \caption{}
        \label{fig:BBH}
    \end{subfigure}
    
    \caption{Same as Fig.~\ref{fig:contour_SHM_LMC}, but using only the full efficiency window 0--650~keV, and assuming (a) the $\text{LMC}_{y=1\%}$ model vs the $\text{LMC}(1100)_{y=10\%}$ model; (b) the boosted DM distribution due to the binary black holes in the galaxy, as well as a hypothetical 1000 times larger flux from this effect (in dashed)~\cite{Acevedo:2026xol}.}
    \label{fig:extended_contour}
\end{figure*}

\subsection{Statistical analysis}
\label{sec:statistics}
To find the $90\%$ confidence contour in the $(\sxn,\delta)$ parameter space (at fixed $m_\chi$) consistent with the observed event, we use the extended maximum likelihood~\cite{Fan:2026kxx},
\begin{align}
    \mathcal{L}(\sxn,\delta) \equiv e^{-\Nsignal(\sxn,\delta)} 
    \left. \frac{d\Nsignal(\sxn,\delta)}{dE_R} \right|_{E_R = E_{\rm obs}}~.
\end{align}
The contour admits a simple analytic expression (see Appendix~\ref{app:analytic_contour}). Note that this method measures only the relative likelihood compared to the best-fit point, but not the absolute goodness-of-fit.

In experiment projection plots, we will also show the constraints set by LZ. To set $90\%$ confidence limit, we use a simple Poisson approximation\footnote{This method roughly reproduces Fig.~S7 of Ref.~\cite{LZ:2026axp}.}: Assume the number of signal events occurring anywhere in the relevant energy window is given by a Poisson distribution. Let $(\sigma_{\chi n0},\delta_0)$ denote the set of parameters that yields 3.89 expected events given the relevant exposure. If only 0 or 1 event were actually observed, then the region $(\sxn>\sigma_{\chi n0},\delta_0)$ can be ruled out at 90\% confidence.

\subsection{Parameter space for $\mx=1.1$~TeV}

\begin{figure*}[t!]
    \centering
    
    \begin{subfigure}{0.49\textwidth}
        \centering
        \includegraphics[width=\textwidth]{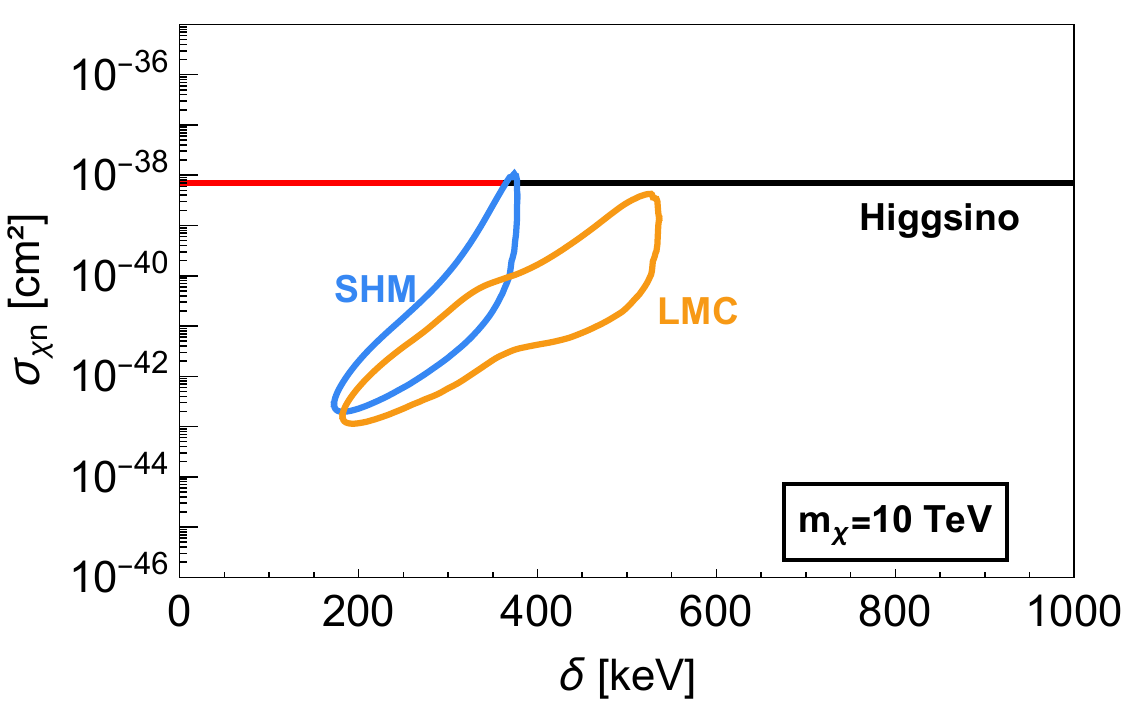}
        \caption{}
        \label{fig:contour_SHM_LMC_10TeV}
    \end{subfigure}
    \hfill
    \begin{subfigure}{0.49\textwidth}
        \centering
        \includegraphics[width=\textwidth]{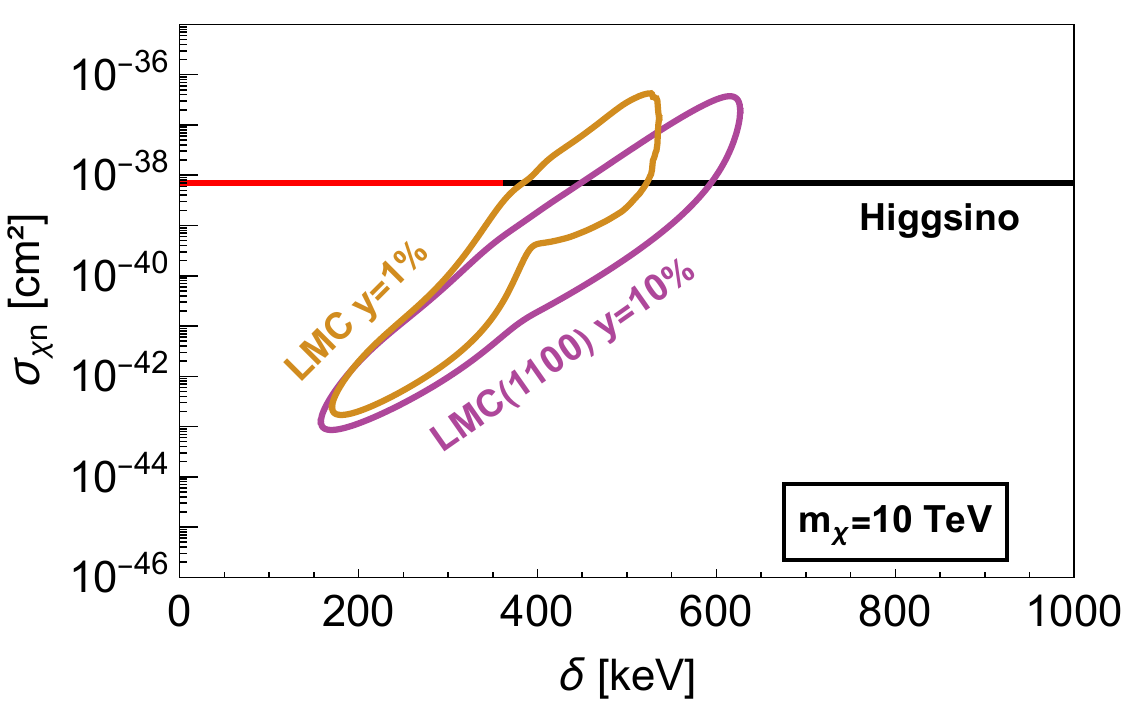}
        \caption{}
        \label{fig:contour_LMC_extend_10TeV}
    \end{subfigure}
    
    \caption{Same as Fig.~\ref{fig:extended_contour}, but for $\mx=10$~TeV. The assumed velocity distributions are (a) the SHM vs the LMC model; (b) the $\text{LMC}_{y=1\%}$ model vs the $\text{LMC}(1100)_{y=10\%}$ model. The solar neutrino constraint on Higgsino (red line) at 10 TeV is $\delta \leq 365$~keV~\cite{Luengas:2026tlf}.}
    \label{fig:10TeV}
\end{figure*}   

We show the resulting $(\sxn,\delta)$ parameter space with $m_\chi = 1.1$~TeV for both the SHM and the LMC model in Fig.~\ref{fig:contour_SHM_LMC}. Assuming the ROI efficiency in Fig.~\ref{fig:SHM_LMC_ROI}, the LZ event contours in both halo models appear compatible with the Higgsino cross section, for $\delta_{\rm SHM} \sim $~340--360~keV~\cite{Fan:2026kxx} and $\delta_{\rm LMC} \sim $~495--515~keV, although they still lie within the region ruled out by the solar neutrino constraint $\delta<566$~keV~\cite{Pospelov:2026ewn}. But once the full efficiency window is taken into account in Fig.~\ref{fig:SHM_LMC_full}, the fact that no events were observed at higher recoil energy shifts the contour regions toward smaller cross sections away from the Higgsino line. Smaller values of $\delta \lesssim 200$~keV are also incompatible, because they would predict more events at low $E_R$, but no events below $\Eobs = 248$~keV were observed either.\footnote{Notice that taking the full efficiency window instead of ROI appears to accommodate intermediate values of $\delta \gtrsim 200$~keV, even though the full efficiency window strictly incorporates higher recoil energy. This is in fact a peculiar feature of the relative likelihood analysis, where the contours are the region of parameter space with likelihood near the best-fit point; it says nothing about the absolute goodness-of-fit. The regions inside the full efficiency contours but not the ROI contours in fact have equal absolute goodness-of-fit in the two efficiency scenarios. We have verified that in all cases of interest here, the best-fit points do fit the LZ event reasonably well, despite these pathological features intrinsic to any relative likelihood analysis.}

 In Fig.~\ref{fig:contour_LMC_extended}, we show the corresponding plot for the $\text{LMC}_{y=1\%}$ and $\text{LMC}(1100)_{y=10\%}$ models (for this and all following plots we only use the full efficiency). For the $\text{LMC}_{y=1\%}$ case, the contour overlaps the Higgsino line between $\delta \sim $~440--510~keV, though within the solar neutrino constraint. Comparing this with Fig.~\ref{fig:SHM_LMC_full}, we see that the fiducial LMC model predicts too many expected events to be compatible with the Higgsino, so the contour prefers smaller cross sections; by reducing the high-velocity DM flux by a factor of 100, the contour is pushed upward to the Higgsino line.
 
 Since a faster velocity tail opens up the kinematic threshold required for accessing larger mass splittings $\delta$ [\Eq{\ref{eq:vmin}}], for $\vmax=1100~\kms$ at $y=10\%$, the contour overlaps the Higgsino line in the region $\delta \sim $~500--600~keV, part of which evades the solar neutrino constraint.  

In Fig.~\ref{fig:BBH}, we show the LZ contour assuming a population of DM boosted by binary black holes (BBH) in the galaxy~\cite{Acevedo:2026xol}. Due to the velocity distribution's abrupt departure from the SHM at very low flux but high velocity (Fig.~\ref{fig:halo}), there are two disjoint regions of the BBH contour: a lower region that is essentially the same as the SHM, and an upper region produced by the BBH. We see that the LZ event is compatible with inelastic DM at large cross section $\sigma \sim 10^{-36} \text{--}10^{-32}~\text{cm}^2$ for $\delta \sim 400 \text{--}1400$~keV. However, to be consistent with the thermal Higgsino beyond the solar neutrino constraint, the flux would need to be larger by 3 orders of magnitude (shown in dashed). It is unclear if it is possible to find such a large flux enhancement, but we briefly mention that a similar boosted flux applies for the single BBH system consisting of Sagittarius A* and an intermediate mass black hole~\cite{Acevedo:2026tur}; but this flux could be enhanced simply if there were more such intermediate mass black holes in the nuclear star cluster.  In any case, we present this example mainly to illustrate the effect of any rare process in the galaxy that boosts a small fraction of the dark matter up to high speeds above the escape velocity.

\subsection{Parameter space for $\mx=10$~TeV}
While the best-motivated model is the 1.1~TeV thermal Higgsino, we have seen that it is only consistent with the LZ event if there is some (plausible) departure from the often-assumed velocity distributions (the SHM and the LMC model). To this end, we also consider other Higgsino masses, whose formation mechanism requires nonstandard cosmology. They are thus less well-motivated compared to the WIMP miracle, but have an easier time fitting to the LZ event in the standard velocity assumptions, as we show below.

In Fig.~\ref{fig:10TeV}, we show the $(\sxn,\delta)$ parameter space for $\mx=10$~TeV as a benchmark non-thermal mass. At this mass, the solar neutrino constraint is less stringent, $\delta \leq 365$~keV~\cite{Luengas:2026tlf}. We see that the tip of the LZ event contour just reaches the Higgsino line at $\delta \sim 365 \text{--}380$~keV in the SHM. In the LMC case, it gets very close at $\delta \sim 530$~keV and it can be shown they are consistent at the 99\% confidence level. Reducing the high-velocity tail flux to $y=1\%$ pushes the contour up to have a large overlap with the Higgsino line in the range $\delta \sim 385\text{--}525$~keV. In the $\text{LMC}(1100)_{y=10\%}$ model, they have a large overlap for $\delta \sim 450 \text{--}595$~keV.

\section{Complementary heavy-element experiments}
\label{sec:experiments}

Even if more data across the three xenon experiments including XENON~\cite{XENON:2023cxc} and PandaX~\cite{PandaX:2024qfu}
confirm the event observed at LZ, a handful of events alone cannot establish whether the
underlying DM model is inelastic or elastic with an additional momentum dependence in the cross
section~\cite{Fitzpatrick:2012ix}. In principle the two are distinguishable since the inelastic model has a minimum recoil energy which sets a dramatic edge. Resolving
this edge, however, requires a large event sample in a single target. Moreover, even within the
inelastic hypothesis, a xenon spectrum constrains only combinations of $\mx$, $\delta$, $\sxn$, and
$\eta$~[\Eq{\ref{eq:dRdER}}], leaving a degeneracy among them.

Heavier target nuclei help on both fronts for a simple kinematic reason. The largest splitting $\delta_{\max}$ grows linearly with the target mass. If the
xenon rate is small because $\delta$ lies close to xenon's kinematic limit, then a heavier target,
for which the same $\delta$ sits well inside the accessible window, samples a much larger fraction
of the DM velocity distribution and yields enhanced rates. This could be accessible at heavy-element experiments. The recoil spectrum for different nuclei can also help break the degeneracies present in a single nuclear target.

Testing the inelastic DM interpretation of the LZ event therefore calls for complementary experiments
with heavier targets. We show that two such experiments proposed by some of
us~\cite{Graham:2026ivn,EnhancingHiggsinoDM_2024}, originally designed to probe mass splittings
beyond the reach of xenon, are also well suited to checking the LZ event and breaking this
degeneracy. Their main qualitative advantages are as follows:
\begin{enumerate}
    \item Heavy-element paleodetectors~\cite{Graham:2026ivn} have the advantage of easily covering different target nuclei, simply by finding minerals made of different elements.\footnote{The method of Ref.~\cite{Graham:2026ivn} generalizes to any element, not just Pb.} The DM-induced track length spectrum scales with the target nuclear mass in a specific way that would give clear evidence of inelastic DM. We will show that just two elements would in principle allow for a determination of $\mx$ and $\delta$. Additionally, paleodetectors' unique sensitivity to the history of the DM velocity distribution can test its LMC origin.
    \item Luminous detection with heavy-element supplements~\cite{EnhancingHiggsinoDM_2024} would immediately reveal the inelastic nature of DM, that it couples to the photon, and directly measure the mass splitting $\delta$. We would also learn if the signal is from a Higgsino since its decay rate is calculable.
\end{enumerate} 

In what follows, we briefly summarize each of these ideas and show their reach into the $(\sxn,\delta)$ parameter space compatible with the observed LZ event.

\begin{figure*}[t!]
    \centering
    
    \begin{subfigure}{0.49\textwidth}
        \centering
        \includegraphics[width=\textwidth]{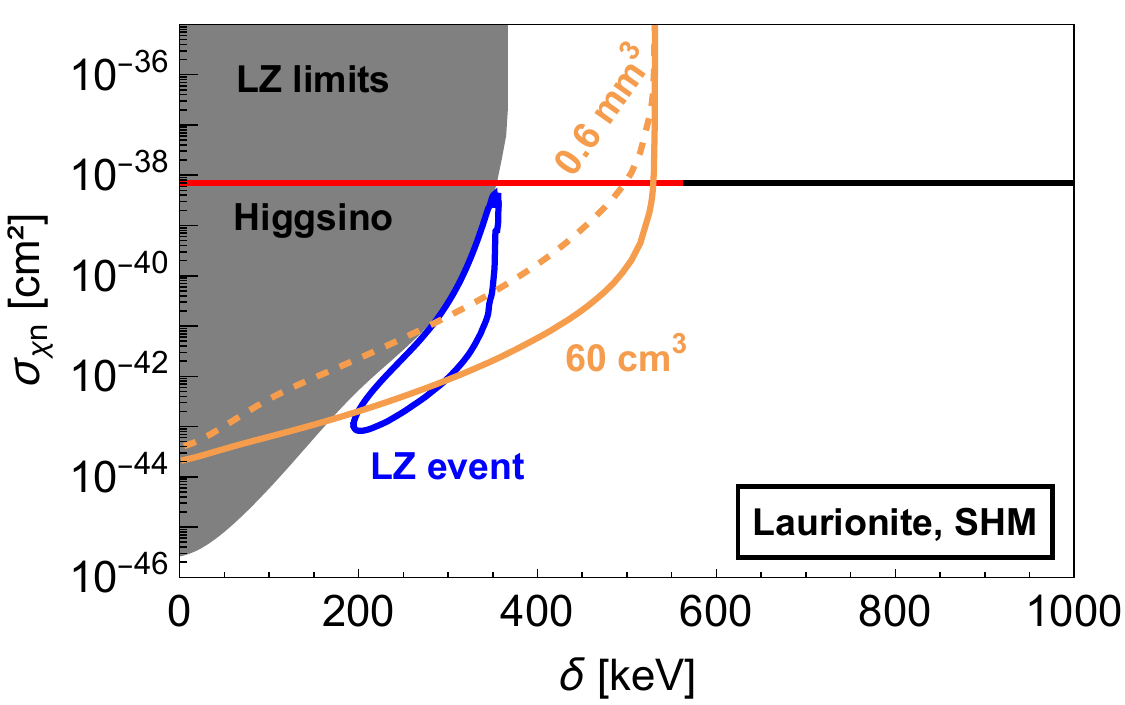}
        \caption{}
        \label{fig:paleoSHM}
    \end{subfigure}
    \hfill
    \begin{subfigure}{0.49\textwidth}
        \centering
        \includegraphics[width=\textwidth]{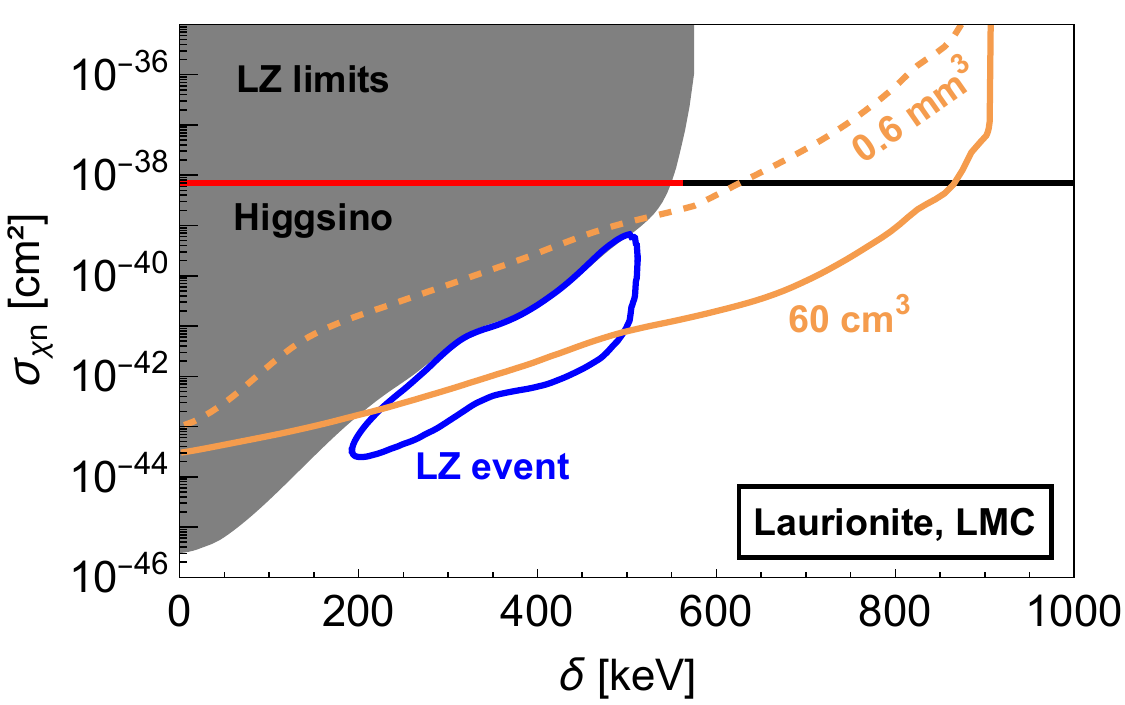}
        \caption{}
        \label{fig:paleoLMC}
    \end{subfigure}
    
    \begin{subfigure}{0.49\textwidth}
        \centering
        \includegraphics[width=\textwidth]{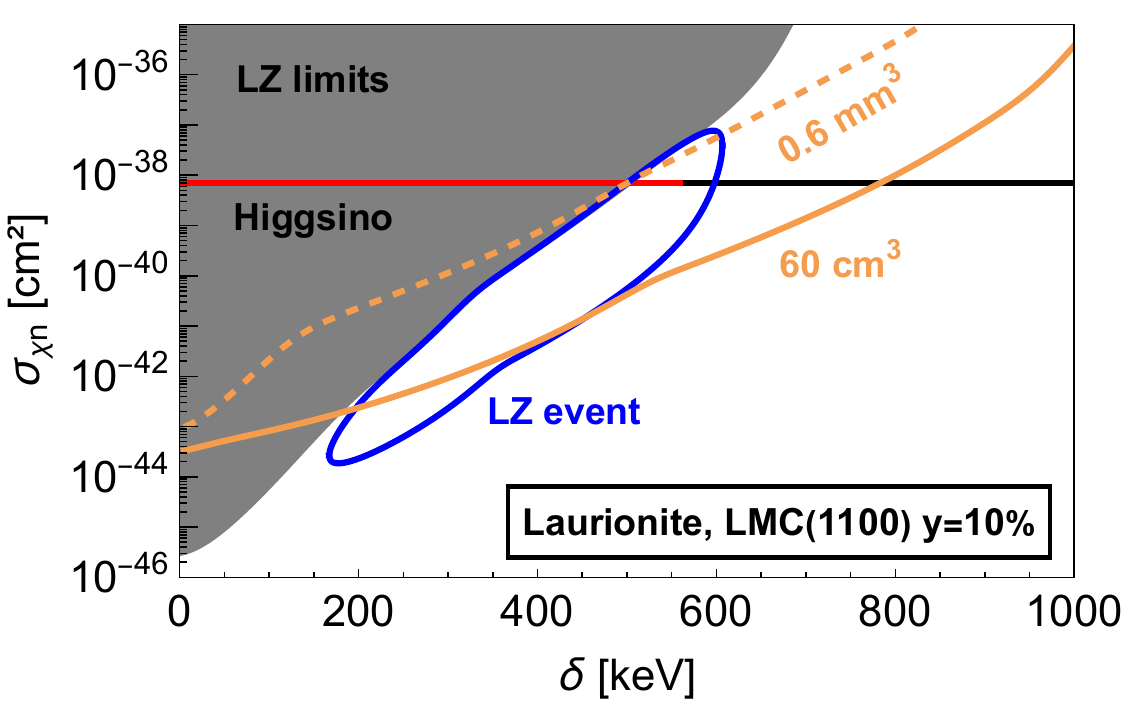}
        \caption{}
        \label{fig:paleo1100_short10}
    \end{subfigure}
    \hfill
    \begin{subfigure}{0.49\textwidth}
        \centering
        \includegraphics[width=\textwidth]{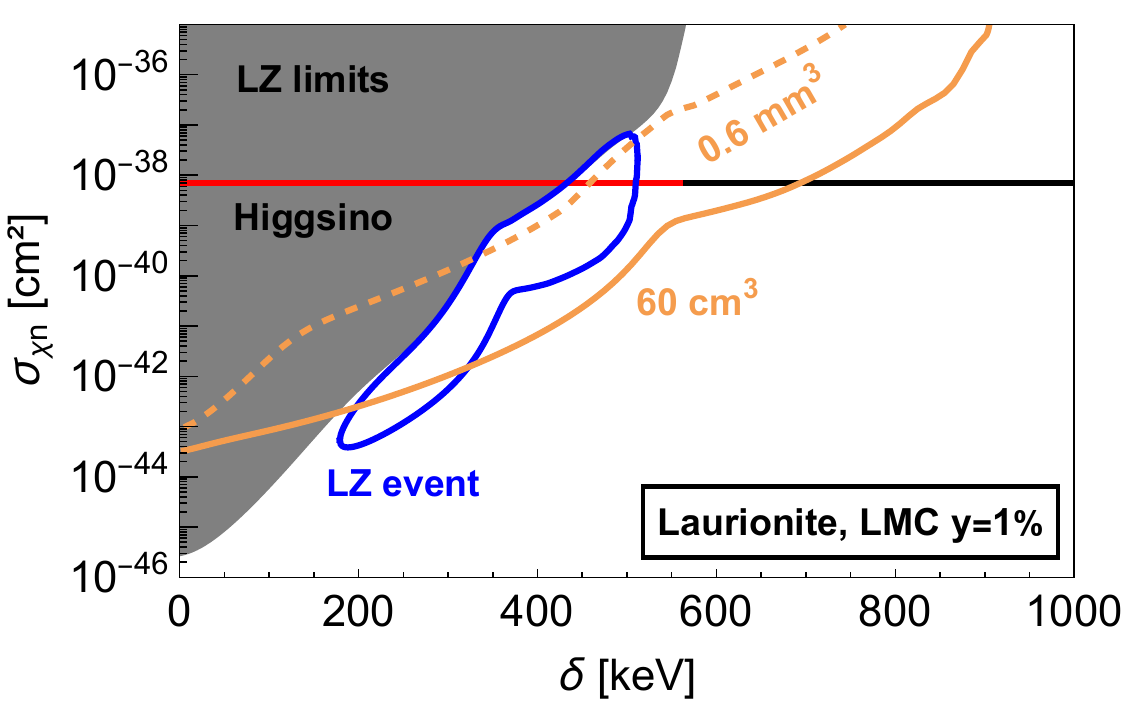}
        \caption{}
        \label{fig:paleo_short100}
    \end{subfigure}
    
    \caption{LZ contours from   Fig.~\ref{fig:contour_SHM_LMC} and Fig.~\ref{fig:extended_contour}, overlaid with the projected sensitivity of Laurionite (PbClOH) paleodetectors, for both a $60~\text{cm}^3$ and a $0.6~\text{mm}^3$ sample, with resolution $\sigma_x=15$~nm relevant for Small Angle X-ray scattering (SAXs). The assumed velocity distributions are (a) the SHM; (b) the LMC model, assuming a linear evolution between pericenter and the present day~\cite{Graham:2026ivn}; (c) the $\text{LMC}(1100)_{y=10\%}$ model; (d) the $\text{LMC}_{y=1\%}$  model (with linear time evolution). The exposure time is taken to be 1~Gyr for the SHM, and 50~Myr for all LMC models. The $90\%$ confidence limit set by LZ is shaded in gray.}
    \label{fig:paleo}
\end{figure*}

\begin{figure*}[t!]
    \centering
    
    \begin{subfigure}{0.49\textwidth}
        \centering
        \includegraphics[width=\textwidth]{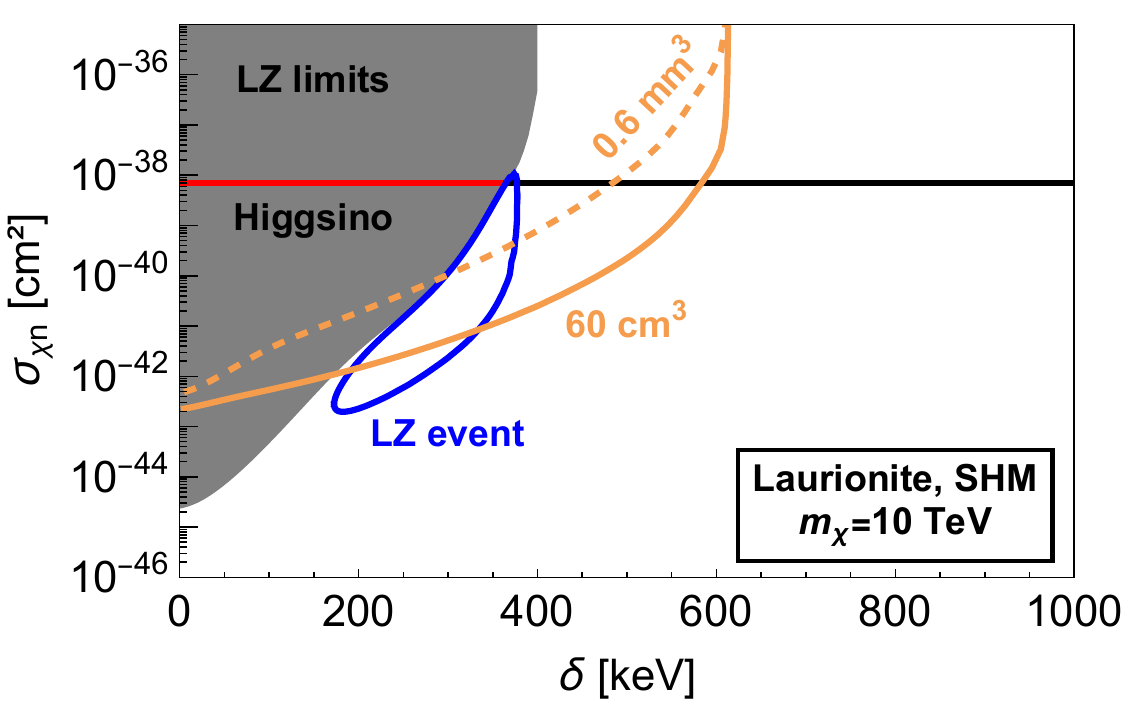}
        \caption{}
        \label{fig:paleoSHM_10TeV}
    \end{subfigure}
    \hfill
    \begin{subfigure}{0.49\textwidth}
        \centering
        \includegraphics[width=\textwidth]{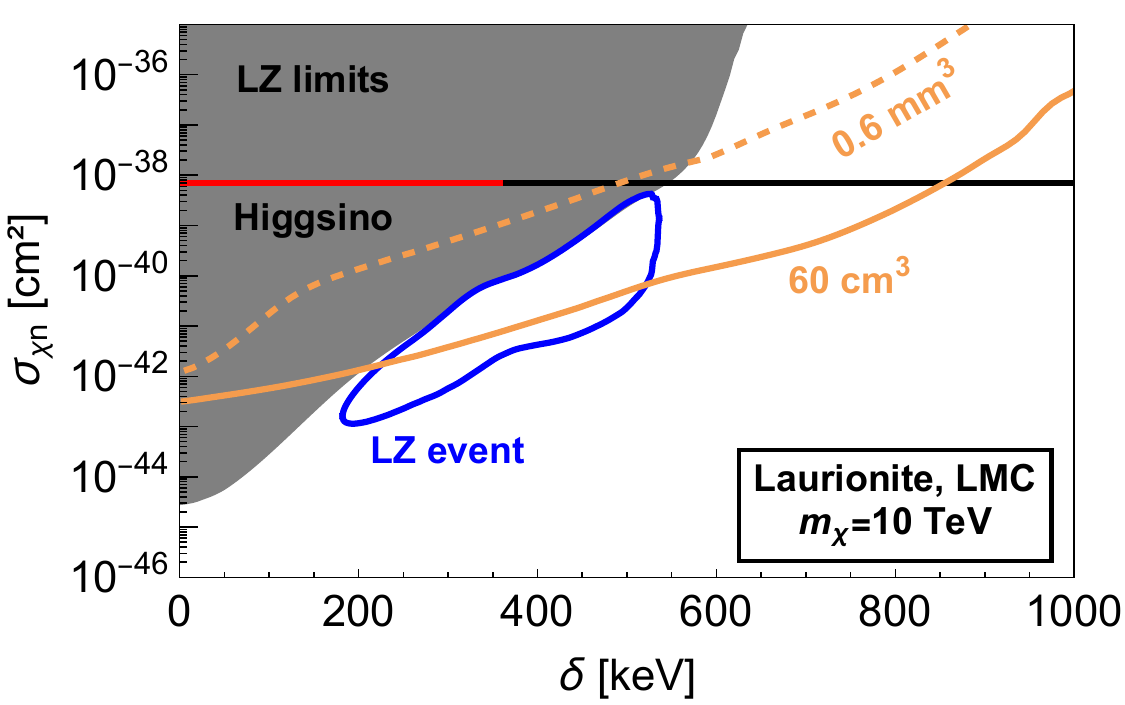}
        \caption{}
        \label{fig:paleoLMC_10TeV}
    \end{subfigure}
    
    \begin{subfigure}{0.49\textwidth}
        \centering
        \includegraphics[width=\textwidth]{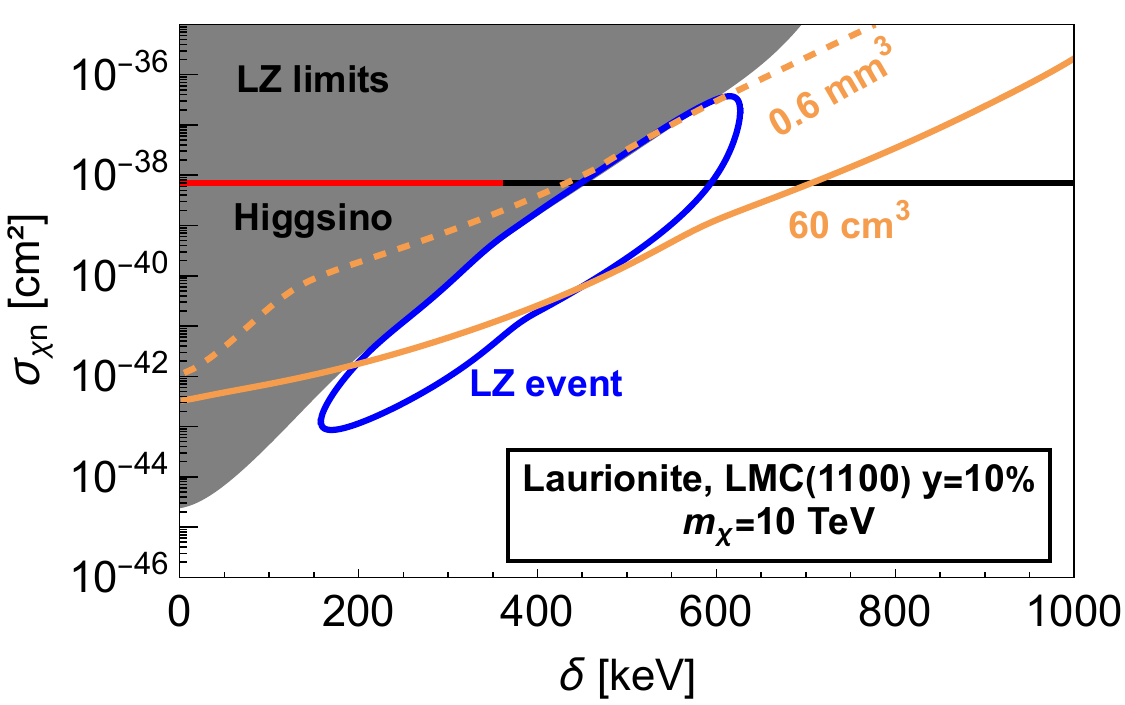}
        \caption{}
        \label{fig:paleo1100_short10_10TeV}
    \end{subfigure}
    \hfill
    \begin{subfigure}{0.49\textwidth}
        \centering
        \includegraphics[width=\textwidth]{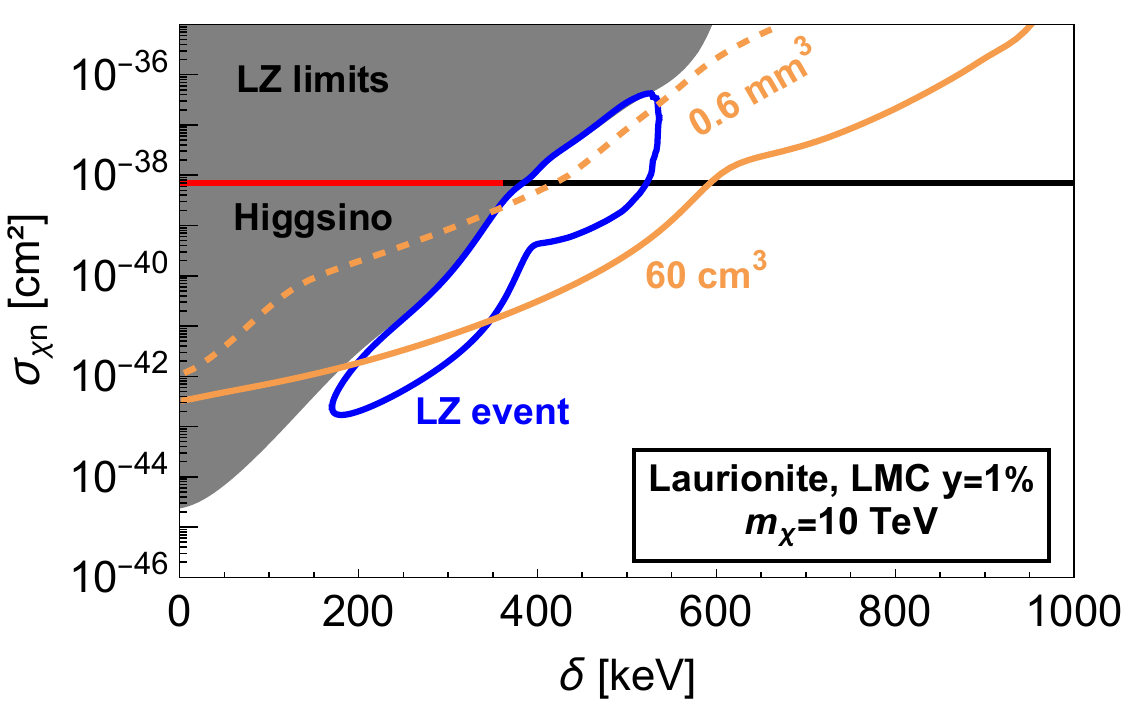}
        \caption{}
        \label{fig:paleo_short100_10TeV}
    \end{subfigure}
    
    \caption{Same as Fig.~\ref{fig:paleo}, but for $\mx=10$~TeV.}
    \label{fig:paleo_10TeV}
\end{figure*}

\subsection{Heavy-element paleodetectors}

The paleodetector program~\cite{paleodetectors_2018,paleodetectors_details_2018,paleodetectors_spectral_2018,paleodetectors_new_2021,paleodetectors_refine_2025,Theodosopoulos:2026ehn,Elykov:2026ojl} aims to search for the persistent damage tracks left in ancient minerals by WIMP-nucleus scattering, taking advantage of the long exposure time (up to $\sim$~Gyr) accumulated in certain rocks. A nuclear recoil with energy $E_R$ creates a damage track with length $x(E_R)$, given by the material's stopping power $dE/dx$. The differential recoil rate $dR/dE_R$ in \Eq{\ref{eq:dRdER}} can then be converted to a track length spectrum $dR/dx$, which spans roughly the $x \sim $~1--300 nm range for a TeV mass WIMP~\cite{Graham:2026ivn,paleodetectors_2018}.

Materials known as solid-state track detectors (SSTDs)~\cite{Fleischer:1964,Fleischer383,Fleischer:1965yv,GUO2012233} can retain these damage tracks over geological timescales, and their three-dimensional reconstruction could be achieved via methods such as Small Angle X-ray scattering (SAXs) tomography~\cite{Schaff:2015}. Candidate minerals need to be sourced from $\sim 5$~km underground to suppress cosmogenic neutron backgrounds, and must be sufficiently radiopure to mitigate radioactive backgrounds. 

However, conventional paleodetectors use minerals whose heaviest element is iron ($A_{\rm Fe}=56$)~\cite{Theodosopoulos:2026ehn}, which is lighter than xenon ($A_{\rm Xe}=131$). This severely limits their sensitivity to inelastic DM; in particular, they are insensitive to the region of $(\sxn,\delta)$ parameter space consistent with the LZ event.

In Ref.~\cite{Graham:2026ivn}, we proposed heavy-element paleodetectors for Higgsino and general inelastic dark matter, showing that minerals containing heavy elements such as lead ($A_{\rm Pb}=207$) provide good sensitivity to large $\delta$. We identified brine precipitation from deep geothermal aquifers as a possible geological formation mechanism that naturally produces Pb-bearing minerals with uranium concentrations as low as $C^{238} \lesssim 3\times 10^{-12}~\gpg$~\cite{Kraemer_lowU_1981,Kraemer_lowU_1986,geopressure_review_1992}. For example, we identified Laurionite (PbClOH) as one promising candidate mineral; it frequently forms in brine precipitates, and its hydrogen content suppresses fast neutron backgrounds.

In Fig.~\ref{fig:paleo}, we show the projected sensitivity of Laurionite, assuming the spatial resolution ($\sigma_x \approx 15$~nm~\cite{SAX_resolution_2014}) and processable volume ($60~\text{cm}^3$) of SAXs read-out~\cite{Schaff:2015}.\footnote{The calculation follows that of Ref.~\cite{Graham:2026ivn}, where we require at least 5 signal events, and a signal-to-noise ratio of SNR=3 against the irreducible backgrounds from astrophysical neutrinos, fast neutrons, and single-$\alpha$ events.} In Fig.~\ref{fig:paleoSHM}, we assume the SHM and an exposure time of 1~Gyr. The projected sensitivity covers the vast majority of the LZ event contour. 

We also show the reach of a $10^5$ times smaller processing volume ($0.6~\text{mm}^3$), which already covers part of the LZ event contour. Note that if only a small volume target were available, the Helium Ion Beam Microscopy (HIBM)~\cite{Hill:2012} read-out can process up to $6~\text{mm}^3$ with a better resolution of $\sigma_x \approx 1$~nm (not shown in figure). Here, we are showing the same resolution as SAXs, such that it truly represents what could be accomplished in the initial stage of an experiment, i.e., $10^{5}$ times shorter scanning time than the full run. As pointed out in Ref.~\cite{Graham:2026ivn}, the large cross sections allowed for the inelastic DM can reduce the scanning volume needed to probe new parameter space by up to a factor of a million compared with the conventional paleodetector program targeting elastic DM.

For the LMC scenarios, we take into account the unique sensitivity of paleodetectors to the history of the DM velocity distribution. The LMC reached pericenter only 50~Myr ago, and we conservatively assume the velocity distribution was just the SHM up until that point.\footnote{\label{ft:LMC_age}This is a robust assumption because the enhancement also relies on the relative motion of the Sun and the LMC. Since the rotation period of the Sun around the MW is $\sim 200$~Myr~\cite{Feng_2018}, the gain from the almost head-on collision was  lost even just $100$~Myr ago as the Sun was moving in the opposite direction.} Then, a 50~Myr old mineral would maximally take advantage of the high-velocity DM available, and any older mineral would be strictly worse for large $\delta$, since it was only accumulating backgrounds but not signal up until that point; in particular, we find this to be the case for the LZ event contour region. So we assume the optimal mineral age of 50~Myr when considering all variants of the LMC model. Furthermore, since the velocity distribution at pericenter was moderately faster than present day (see Fig.~2 in Ref.~\cite{Graham:2026ivn}; the LMC distribution shown in Fig.~\ref{fig:halo} is for the present-day only), we assume the LMC halo integral evolved linearly in the intervening time, following the treatment in Ref.~\cite{Graham:2026ivn}.\footnote{Since the $\text{LMC}(1100)_{y=10\%}$ model postulates an extended velocity tail, we assume this halo model has been static for 50~Myr.}

Fig.~\ref{fig:paleoLMC} shows the sensitivity in the LMC model, where the paleodetector continues to cover most of the LZ event contour, and covers the Higgsino up to $\delta = 865$~keV, although the Higgsino still cannot explain the LZ event in this scenario.

In the $\text{LMC}(1100)_{y=10\%}$ and $\text{LMC}_{y=1\%}$ models in Figs.~\ref{fig:paleo1100_short10} and \ref{fig:paleo_short100}, respectively, all overlap of the LZ contours with the Higgsino line is covered by the paleodetectors, whose full Higgsino reach goes up to $\delta \sim $~780~keV and $\delta \sim 695$~keV respectively.  Further, almost all of the general inelastic DM parameter space that fits the LZ event is also covered by this paleodetection.

We also show the corresponding reach for $\mx=10$~TeV in Fig.~\ref{fig:paleo_10TeV}. Here, the paleodetector continues to be an excellent probe, covering most of the LZ event contour in all scenarios, and in particular, covers all overlap regions between the contour and the Higgsino line.

In addition to confirming the LZ event, unique features of paleodetectors give them several distinctive advantages in testing the inelastic DM hypothesis. With only a handful of events from xenon detectors, one cannot determine whether the signal comes from inelastic DM or some other non-standard DM model. But if different target nuclei are available, the inelastic nature becomes easily testable: the spectrum depends sensitively on the available velocity phase space, which depends on the target nuclear mass via $\vmin(E_R)$~[\Eq{\ref{eq:vmin}}], and nuclear mass below the threshold $\mu_A <2 \delta/\vmax^2$ will not see any signal. In general, it is very hard to just build new direct detection experiments with other elements, as they are all designed differently. But for paleodetectors, we just need to search for new mineral types; the same read-out method can be applied to different elements.

In fact, with just two different elements, it is possible in principle to determine the dark matter mass $\mx$ and splitting $\delta$ independently. To see this, consider the track length spectrum $dR/dx$ obtained from a paleodetector. Since the material's $dE/dx$ can be experimentally determined, we can deduce the damage track length as a function of recoil energy, $x(E_R)$, which allows us to reconstruct the energy spectrum $dR/dE_R$ in \Eq{\ref{eq:dRdER}}. Since $F_A^2(E_R)$ is known for a given element, we can divide it to get the spectrum
\begin{equation}
s(E_R) \equiv \frac{1}{F_A^2(E_R)}\frac{dR}{dE_R}
=
\frac{\rhoDM \sxn}
{2 m_\chi \mu_n^2} w_A Q_A^2
 \times
\eta\!\left(v_{\min}(E_R)\right) ~,
\end{equation}
where the only dependence on $E_R$ arises from the term $\eta\!\left(v_{\min}(E_R)\right)$. Since $\eta$ is a strictly decreasing function of $v_{\min}$ (at least in all the cases we consider), the maximum of $s(E_R)$ occurs at the minimum of $v_{\min}(E_R)$. But this is just the well-known ``apex point" on the $E_R$ vs $\vmin$ contour (see Fig.~1 of Ref.~\cite{Graham:2026ivn}), $v_{\min}^{\rm apex} = \sqrt{2\delta/\mu_A}$, which occurs at the apex recoil energy 
\begin{align}
    E_R^{\rm apex}(A) = \frac{\mu_A}{m_A} \delta = \frac{\mx \delta}{\mx + m_A} ~.
\end{align}
So by locating the peak of the reconstructed spectrum $s(E_R)$, we can find the value of $E_R^{\rm apex}(A)$. This is an equation with two unknowns $\mx$ and $\delta$. Suppose we have two paleodetectors made of two different heavy elements, $A_1$ and $A_2$. Then measuring $E_R^{\rm apex}(A_1)$ and $E_R^{\rm apex}(A_2)$ allows us to deduce both $\mx$ and $\delta$. This method works best when $\mx$ is not too heavy; when $\mx \gg m_A$ the $\mx$ cancels between the numerator and denominator. 

Another advantage is that paleodetectors' unique sensitivity to the history of the DM velocity distribution can test whether something like this LMC scenario is the source of the high-velocity tail responsible for this event. Consider a sequence of heavy-element mineral samples whose ages $t$ span both much younger and much older than $t_{\rm LMC} \sim 50$~Myr. The signal accumulates as $S\propto t$ for $t<t_{\rm LMC}$ and saturates for $t>t_{\rm LMC}$, whereas the background grows as $B\propto t$ for all ages. The signal-to-noise ratio $\text{SNR}(t)\propto S/\sqrt{B}$ therefore scales as $\sqrt{t}$ until it peaks at $t \sim t_{\rm LMC}$, then falls as $1/\sqrt{t}$. We expect $t_{\rm LMC}\sim 50$--$100$~Myr, with the precise value depending on detailed modeling of the LMC-MW encounter (see also footnote~\ref{ft:LMC_age}).  An observation of such a distinctive time-dependence would be evidence for the LMC-produced high-velocity tail, and it would also strongly favor the thermal Higgsino, whose explanation of the LZ event requires DM speeds beyond the cutoff of the SHM.

\begin{figure*}[t!]
    \centering
    
    \begin{subfigure}{0.49\textwidth}
        \centering
        \includegraphics[width=\textwidth]{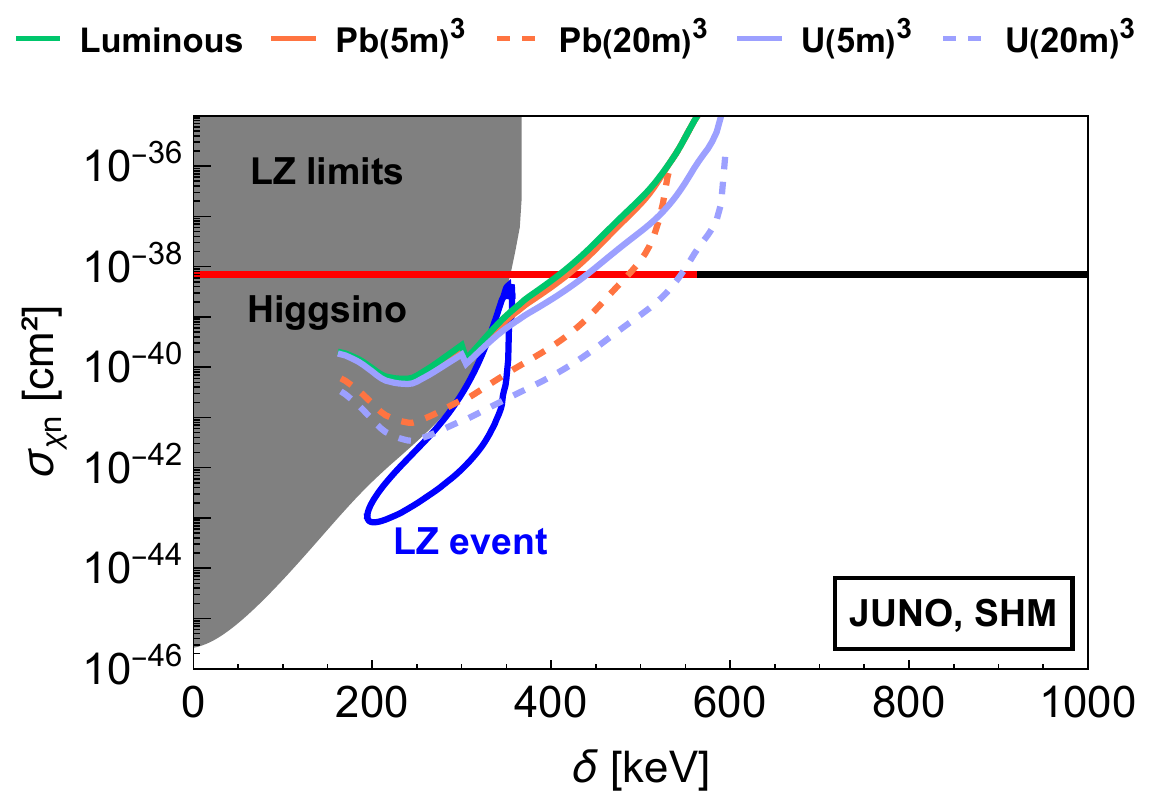}
        \caption{}
        \label{fig:junoSHM}
    \end{subfigure}
    \hfill
    \begin{subfigure}{0.49\textwidth}
        \centering
        \includegraphics[width=\textwidth]{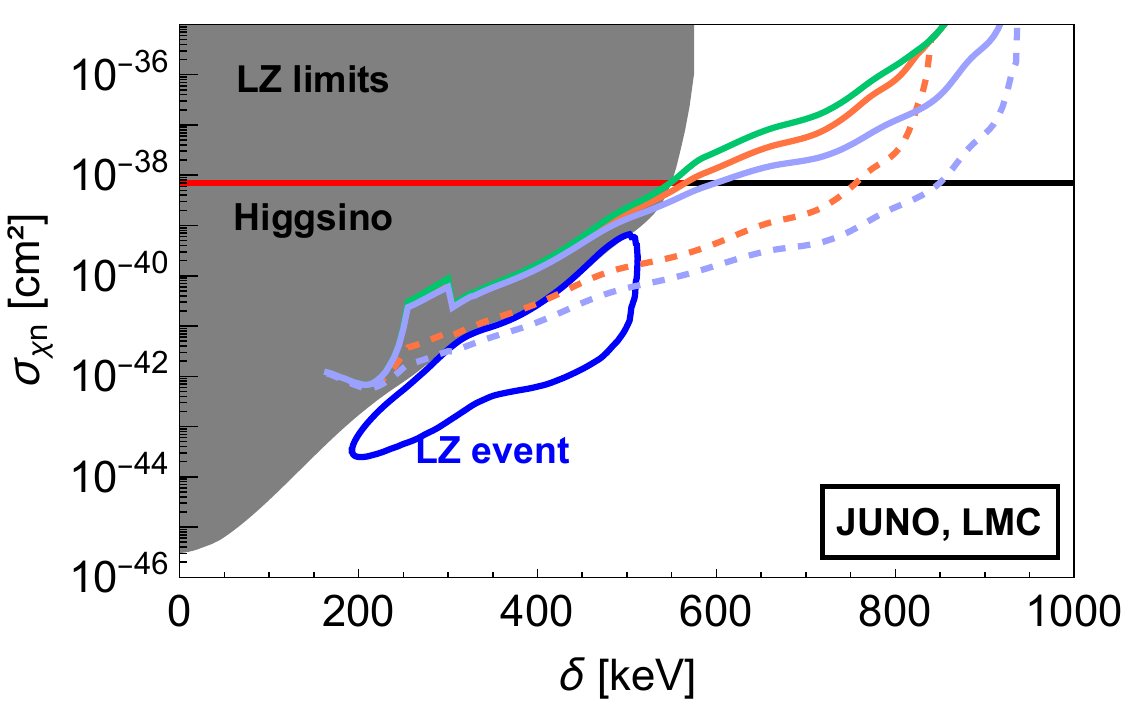}
        \caption{}
        \label{fig:junoLMC}
    \end{subfigure}
    
    \begin{subfigure}{0.49\textwidth}
        \centering
        \includegraphics[width=\textwidth]{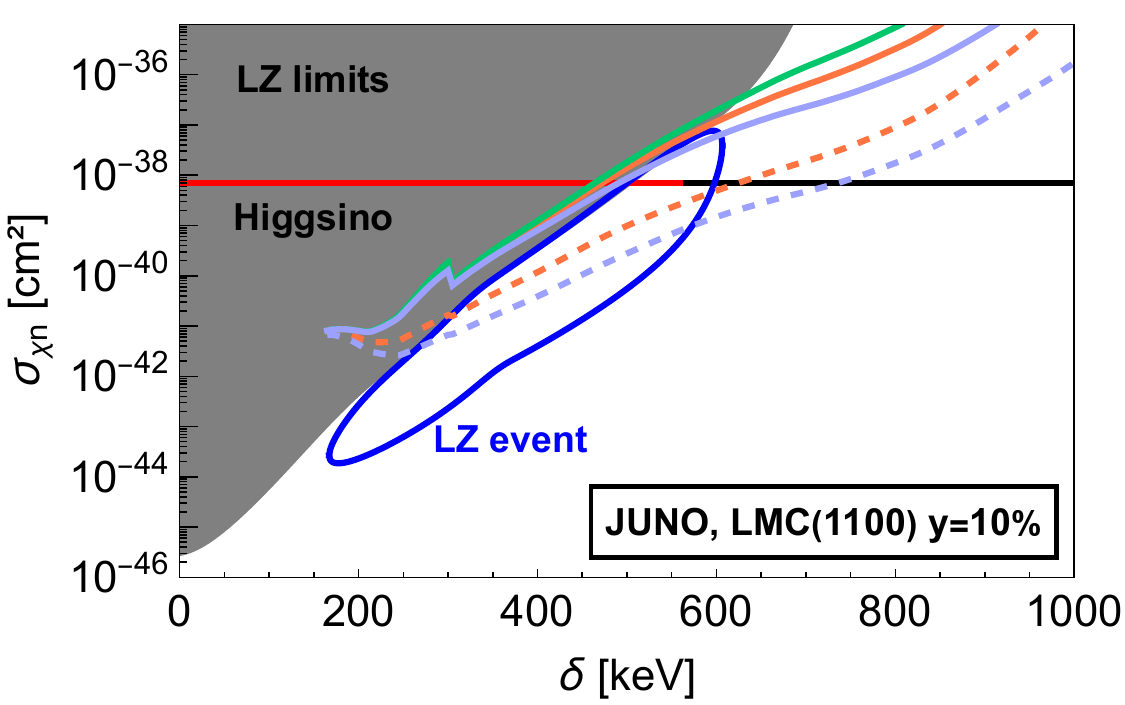}
        \caption{}
        \label{fig:juno1100_short10}
    \end{subfigure}
    \hfill
    \begin{subfigure}{0.49\textwidth}
        \centering
        \includegraphics[width=\textwidth]{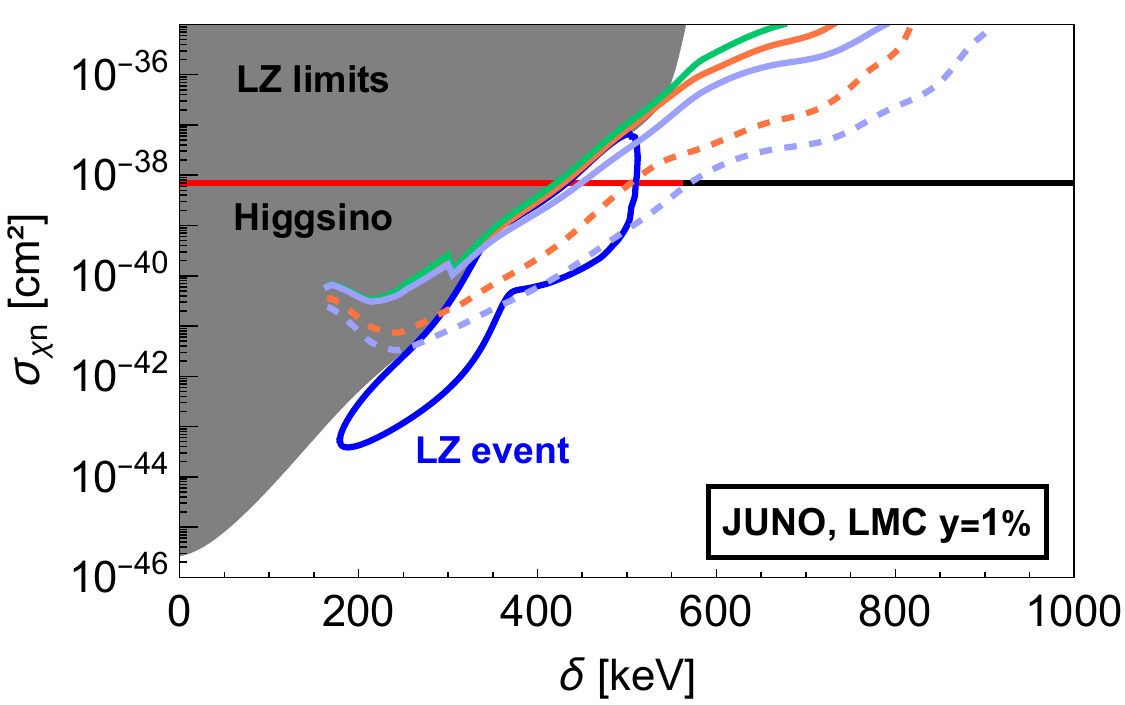}
        \caption{}
        \label{fig:juno_short100}
    \end{subfigure}
    
    \caption{The same LZ contours and LZ limits as in Fig.~\ref{fig:paleo}, but overlaid with the projected sensitivity of JUNO~\cite{JUNO_Conceptual} assuming $\sim 10$~year exposure time, for both its luminous signal from natural elements in the Earth (green), as well as supplementing with various volumes of Pb (orange) or $^{238}\text{U}$ (purple).}
    \label{fig:juno}
\end{figure*}

\subsection{Luminous detection with heavy-element supplements}

The ``luminous dark matter" detection technique takes advantage of  the fact that the excited Higgsino state undergoes radiative decay, $\chi' \to \chi+\gamma$~\cite{Feldstein:2010su,Eby:2019mgs,EnhancingHiggsinoDM_2024}. An incoming DM particle $\chi$ upscatters against a heavy nucleus in the Earth into the excited state $\chi'$, continues propagating until it subsequently decays, emitting a monochromatic photon with energy $E_{\gamma} = \delta$ that can be detected by a large-volume underground detector, such as the neutrino experiments  JUNO~\cite{JUNO_Conceptual}, SNO+~\cite{SNO+}, or KamLAND~\cite{Kamland}. For an excited state with detector-frame speed $v$, the Higgsino decay length  is~\cite{Krall_2018,HABER1989267}
\begin{align} \label{eq:decay_length}
\ell^{}_{\chi'} = 7.46~\mathrm{km}
\left(\frac{v}{400~\mathrm{km/s}}\right)
\left(\frac{400~\mathrm{keV}}{\delta}\right)^3 \left(\frac{\mx}{1.1~\text{TeV}} \right)^2~.
\end{align}
For all scenarios of interest here, this satisfies the hierarchy 
\begin{align} \label{eq:hierarchy}
    R_D \ll \ell^{}_{\chi'} \ll R_\oplus ~,
\end{align}
where $R_D$ is the size of the detector and $R_\oplus$ is the radius of the Earth. This allows the excited state to enter the detector before decaying, and also ensures its decay before leaving the Earth, giving an unsuppressed luminous signal.

Note that this luminous method of detection applies to inelastic models broader than the Higgsinos. This is true so long as the excited state decays back to the DM via photon emission. This is true in a broad class of models such as those with electric and magnetic transition dipole moments~\cite{Feldstein:2010su,Bramante:2016rdh,Eby:2019mgs,Asadi:2026iot}. Our approximation and the plots that follow apply to these models so long as the decay length satisfies Eq.~(\ref{eq:hierarchy}).

Following Ref.~\cite{EnhancingHiggsinoDM_2024}, we model the Earth as an infinite medium of uniform density, and the signal photon rate $\Gamma_{\gamma}$ is given by
\begin{equation}
\Gamma_{\gamma}=\sum_A \Gamma_{\chi A}\,n_A\,V_{\mathrm{eff},A},
    \label{eq:rateeq}
\end{equation}
where $\Gamma_{\chi A}$ is the DM-nucleus scattering rate, $n_A$ is the number density of element $A$, and $V_{\mathrm{eff},A}$ is its effective volume contributing to the signal. In the infinite Earth approximation, each upscattered particle must decay somewhere else in the Earth, resulting in a uniform distribution of photon signal, and consequently the effective volume must equal the detector volume, $V_{\mathrm{eff},A}=V_D$. The resulting line signal can be distinguished from radioactive and solar neutrino backgrounds via a line search. 

In the SHM, the fastest DM originates from the direction of Cygnus, and the photon signal is present during the half of the sidereal day where Cygnus is below the horizon. Due to the opposite direction between the solar motion and the LMC at pericenter approach, the fast DM from the LMC coincidentally also originates from the Cygnus direction~\cite{EnhancingHiggsinoDM_2024}.

As pointed out in Ref.~\cite{EnhancingHiggsinoDM_2024}, this photon signal can be significantly enhanced by supplementing with heavy elements: placing additional volume of Pb or depleted $^{238}\text{U}$ surrounding the detector. This raises the photon rate due to the much higher number density. The signal from the supplement is present throughout the sidereal day. If a signal were detected, it could be double checked by removing the supplement.

We consider the benchmark supplement volumes $V_{\rm supp}=(5~\text{m})^3$ and $(20~\text{m})^3$. Note that there are $\sim 10^6$ tons of depleted uranium worldwide, with no known application at scale apart from our proposal~\cite{depleteduranium}; hence, the limiting factors are likely issues such as the available space near the detector or the cost of excavating additional volume and moving material. We model the supplement as a spherical shell surrounding the detector, which gives effective volume
\begin{align} \label{eq:Vsupp}
V_{\rm eff}^{\rm supp}(\ell_{\chi'})
=
V_D\,
\frac{(3V_{\rm supp}/4\pi+R_D^3)^{1/3}-R_D}{\ell_{\chi'}}~,
\end{align}
where the signal is suppressed by the decay length, in contrast to the luminous signal (i.e., Earth-only signal) which has a much larger target volume.\footnote{Since the luminous signal does not depend on the precise decay length, its result applies beyond the Higgsino to any inelastic DM model that couples to the photon and whose decay length satisfies the hierarchy~\Eq{\ref{eq:hierarchy}}. On the other hand, the supplement result only generalizes to inelastic DM with the same decay length as the Higgsino [\Eq{\ref{eq:decay_length}}], though rescaling to other decay lengths is trivial.}

A positive signal immediately gives us a lot of important information about dark matter beyond what is possible from a xenon detector. It confirms the inelastic nature of the DM, and furthermore, that it couples to the photon, which is not necessarily true for generic inelastic DM. The value of the photon line $E_{\gamma} = \delta$ immediately tells us the mass splitting. By varying the amount of supplement, we can also deduce the decay length $\ell_{\chi'}$ via \Eq{\ref{eq:Vsupp}}. This would reveal if it is a Higgsino since the Higgsino decay length is calculable [\Eq{\ref{eq:decay_length}}]. In general, it is very difficult to break the degeneracy between $\sxn$ and $\eta$, but a decay length matching that of the Higgsino would avoid this issue.

In Fig.~\ref{fig:juno}, we show the reach of JUNO~\cite{JUNO_Conceptual}, which, at $R_D = 17.5$~m, is the largest neutrino detector with sensitivity to the sub-MeV energy range; the reach for other neutrino detectors is shown in Appendix~\ref{app:other_detectors}. We assume a $\sim 10$~year integration time for all detectors.

In the SHM (Fig.~\ref{fig:junoSHM}), JUNO's luminous signal is already sensitive to a portion of the LZ event contour. With a large volume of supplements, it could probe the upper half of the contour.

In the LMC model (Fig.~\ref{fig:junoLMC}), neither the luminous signal nor the $(5~\text{m})^3$ supplement is sufficient to probe the LZ event contour, though they could reach new Higgsino parameter space $\delta_{\rm Pb} \sim 570$~keV and $\delta_{\rm U} \sim 600$~keV. With $(20~\text{m})^3$ supplement, the upper portion of the LZ event contour could be probed, and they can reach the Higgsino mass splitting $\delta_{\rm Pb} \sim 760$~keV and $\delta_{\rm U}  \sim 850$~keV.

In the $\text{LMC}(1100)_{y=10\%}$ model (Fig.~\ref{fig:juno1100_short10}), the $(20~\text{m})^3$ supplements cover the interesting Higgsino region $566~\text{keV}<\delta<600~\text{keV}$, where the LZ event could be explained by a thermal Higgsino not ruled out by solar neutrino bound; they also have Higgsino reach $\delta_{\rm Pb} \sim 625$~keV and $\delta_{\rm U}  \sim 740$~keV.

In the $\text{LMC}_{y=1\%}$ model (Fig.~\ref{fig:juno_short100}), luminous and small supplement volumes only have little overlap with the LZ event contour, whereas the $(20~\text{m})^3$ supplements cover most of the contour, with Higgsino reach $\delta_{\rm Pb} \sim 505$~keV and $\delta_{\rm U}  \sim 570$~keV.

We also show the JUNO reach for $\mx=10$~TeV in Fig.~\ref{fig:juno_10TeV}. Here, only $(20~\text{m})^3$ supplements can cover upper parts of the LZ event contours.

\section{Conclusion}
\label{sec: conclusion}
The inelastic dark matter hypothesis of the $E_R = 248$~keV event observed by LZ is a promising and well-motivated direction. We have discussed two complementary experiments that can test this hypothesis.  The parts of model space that fit the LZ event depend significantly on the high-velocity tail of the dark matter distribution.  This also affects the observability by complementary experiments. Even changes to this tail that are below the resolutions of existing simulations have significant effects on the dark matter model space.  So we have considered several examples for this tail.

While the best-motivated thermal Higgsino model does not fit the event well in standard halo assumptions, we identify an example velocity distribution that would fit with the Higgsino. This distribution requires changes to the high-velocity tail of the DM distribution at very low flux levels, below the resolution of current simulations. For example, simulations of the LMC indicate that it can give a larger tail of high-velocity DM, even above the galactic escape velocity.
We have shown that a tail which extends up to velocities above $1000~\kms$ could make the thermal Higgsino interpretation viable. This extended velocity tail, while speculative, needs only a tiny flux that is an order of magnitude below the smallest flux reported by current simulations of the LMC. This finding clearly motivates more detailed simulations of the high-velocity tail at better resolution. We also show that a non-thermal Higgsino at 10~TeV fits the event.

In order to test the inelastic DM hypothesis it is useful and possibly necessary to have complementary experiments.  The rough spectrum in a xenon detector could come from other models as well, for example, a momentum-dependent cross section.  Further, xenon experiments alone could not break the various degeneracies between cross section, DM mass, mass splitting, and velocity distribution.  So if we wish to learn more about the dark matter and measure parameters of the underlying model, we will need complementary experiments.

We have demonstrated that two heavy-element experiments proposed by us are well suited for this purpose. Heavy-element paleodetectors~\cite{Graham:2026ivn} naturally cover most of the inelastic DM parameter space consistent with the LZ event. Importantly, they have unique sensitivity to the scattering of DM with other elements heavier than Xe, allowing a probe of the inelastic nature and breaking degeneracies in the underlying model. Further, their sensitivity to the history of the velocity distribution can provide more information on variations in the high-velocity tail, such as could arise if the signal has an LMC origin. An affirmative answer would favor the thermal Higgsino, since the thermal Higgsino explanation of the LZ event is incompatible with the SHM.

Meanwhile, luminous detection with heavy-element supplements~\cite{EnhancingHiggsinoDM_2024} could confirm not only the inelastic nature of the DM but also that it couples to the photon. The observed photon frequency would directly reveal the mass splitting, and varying the supplement would allow for measuring the radiative decay rate, which has a precise prediction in the Higgsino model.

We have identified an actionable path toward testing the inelastic DM interpretation of the LZ event via correlated signals in qualitatively different experiments. More generally, these methods provide powerful probes of inelastic DM even if the LZ event is not dark matter.

\acknowledgments
S.~W. is supported in part by the U.S. Department of Energy Office of Science under Award Number DE-SC0024375, and by the Gordon and Betty Moore Foundation through Grant GBMF13898 to the University of 
Washington.
This work was supported in part by NSF Grant No. PHY-2609853, NSF Grant No. PHY-2515007, the NSF QSEnSE quantum center award, Simons Investigator Award No. 824870, The University of Delaware Research Foundation and the John Templeton Foundation Award No. 63595.

\appendix

\section{Likelihood contour}
\label{app:analytic_contour}
The likelihood contour admits the analytic form,
\begin{align}
    \sxn^{\rm contour}(\delta) = \frac{-1}{\tilde{N}_{\rm signal}(\delta)}W_k\left[ - \frac{p(E_{\rm obs},\delta_{\rm opt})}{p(E_{\rm obs},\delta)} e^{-(1+\Delta \chi^2/2)}\right] ~,
\end{align}
where $\tilde{N}_{\rm signal}=\Nsignal/\sxn$; $p(E_R,\delta)$ is the normalized spectrum,
\begin{align}
    p(E_R,\delta)= \frac{1}{\Nsignal} \frac{d\Nsignal}{dE_R} ~;
\end{align}
$\delta_{\rm opt}$ is the splitting that maximizes $p(E_{\rm obs},\delta)$; $\Delta \chi^2 = 4.6$ is the $\chi^2$ value at $90\%$ confidence for 2 degrees of freedom;
$W_k(x)$ is the Lambert W function, with $k=0$ and $k=-1$ giving the lower and upper contour curves, respectively. 

\section{Detector-frame velocity distribution}
\label{app:detector_frame}
To convert a given velocity distribution from the galactic frame to the detector frame, we boost it by the detector's velocity with respect to the MW,
\begin{align}
    f^{\rm det}(\vbold,t) = f^{\rm gal}(\vbold +\vbold_{\rm det}(t))~.
\end{align}
The detector velocity is taken to be
\begin{align}
    \vbold_{\rm det}(t) = \vbold_\odot + \vbold_\oplus(t) ~,
\end{align}
where the Sun's velocity with respect to the MW in galactic coordinates (where $\hat{x} $ points to the galactic center and $\hat{y}$ is the direction of galactic rotation) is taken to be $\vbold_\odot = (11.1,232.24,7.25)$~km/s~\cite{Smith-Orlik:2023kyl}; the Earth's velocity with respect to the Sun is taken to be $\vbold_\oplus(t) = v_\oplus(\hat{\epsilon}_1 \cos \omega t + \hat{\epsilon}_2 \sin \omega t)$, with $v_\oplus = 29.79$~km/s, $\hat{\epsilon}_1= (0.9940,0.1095,0.003116)$, and $\hat{\epsilon}_2= (-0.05173,0.4945,-0.8677)$~\cite{Lee:2013xxa}. The maximum detector velocity in galactic frame is 
\begin{align}
    v_{\rm det}^{\rm max} = \sqrt{v_\odot^2 + v_\oplus^2 + 2v_\oplus\sqrt{(\vbold_\odot \cdot \hat{\epsilon}_1)^2 + (\vbold_\odot \cdot \hat{\epsilon}_2)^2}} \approx 249~\text{km}/\text{s}~.
\end{align}
Then the maximum DM velocity in the detector frame is
\begin{align}
    \vmax = v^{\rm gal}_{\rm max} + v_{\rm det}^{\rm max} ~.
\end{align}
Finally, we average $f^{\rm det}(\vbold,t)$ over the Earth's annual rotation.

\section{Results for other neutrino detectors}
\label{app:other_detectors}

\begin{figure*}[t]
    \centering
    
    \begin{subfigure}{0.49\textwidth}
        \centering
        \includegraphics[width=\textwidth]{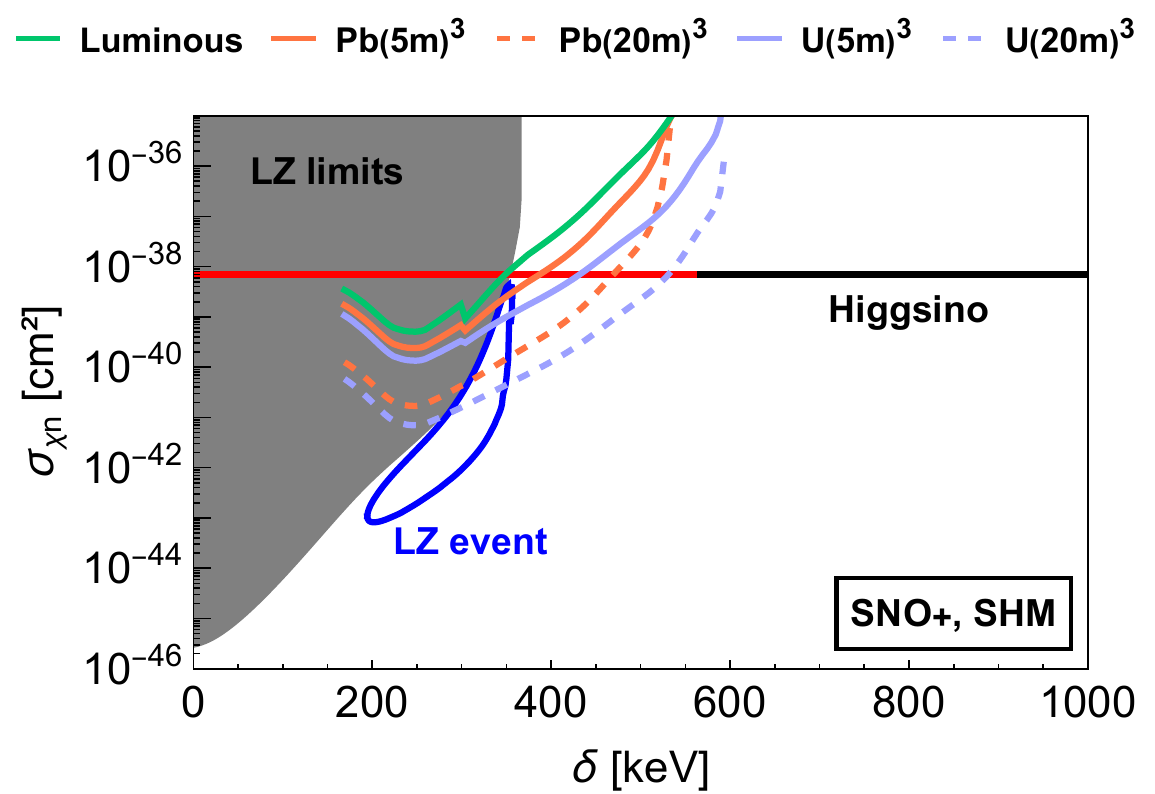}
        \caption{}
        \label{fig:snoSHM}
    \end{subfigure}
    \hfill
    \begin{subfigure}{0.49\textwidth}
        \centering
        \includegraphics[width=\textwidth]{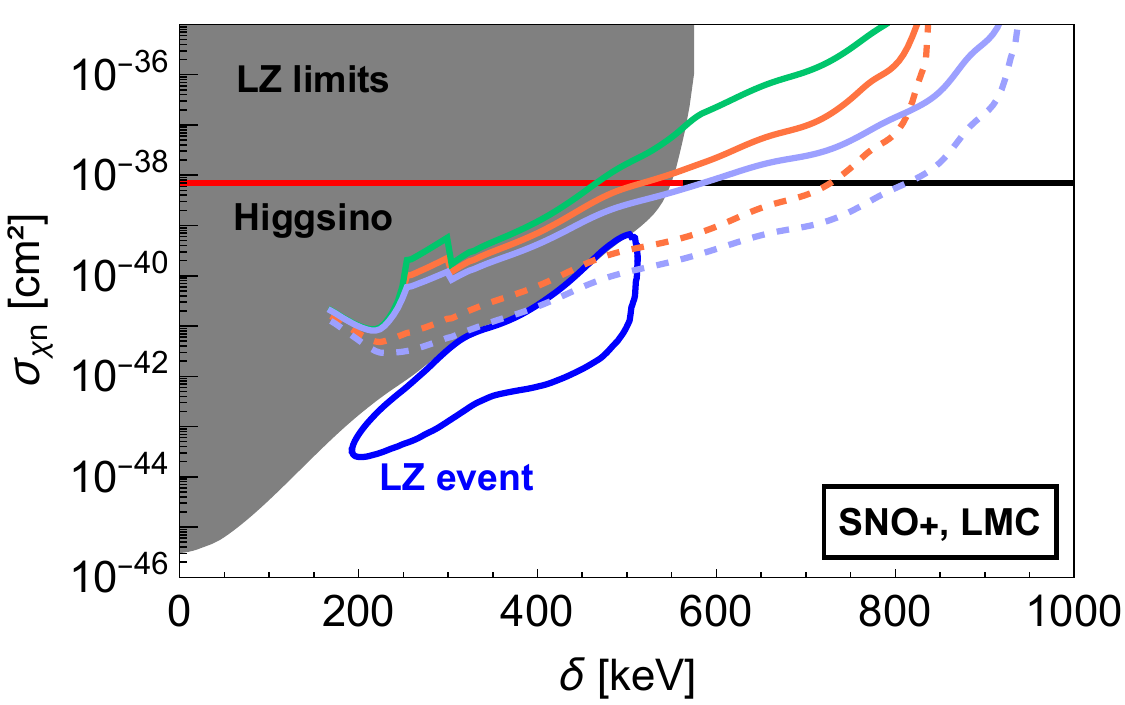}
        \caption{}
        \label{fig:snoLMC}
    \end{subfigure}
    
    \begin{subfigure}{0.49\textwidth}
        \centering
        \includegraphics[width=\textwidth]{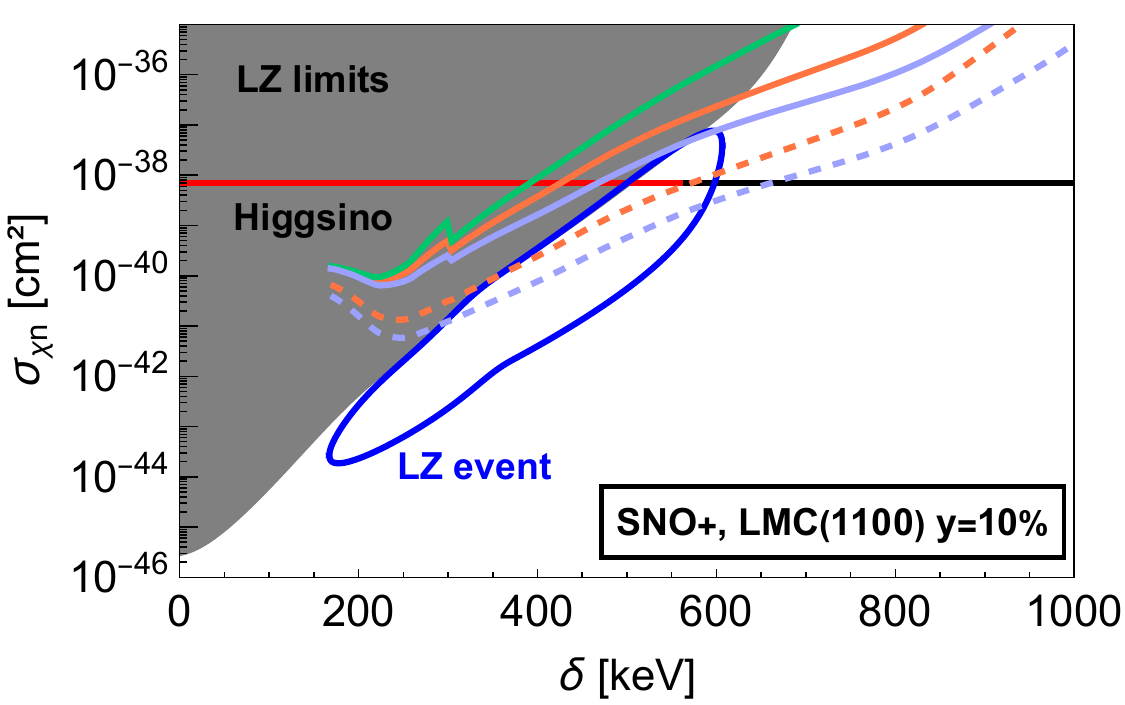}
        \caption{}
        \label{fig:sno1100_short10}
    \end{subfigure}
    \hfill
    \begin{subfigure}{0.49\textwidth}
        \centering
        \includegraphics[width=\textwidth]{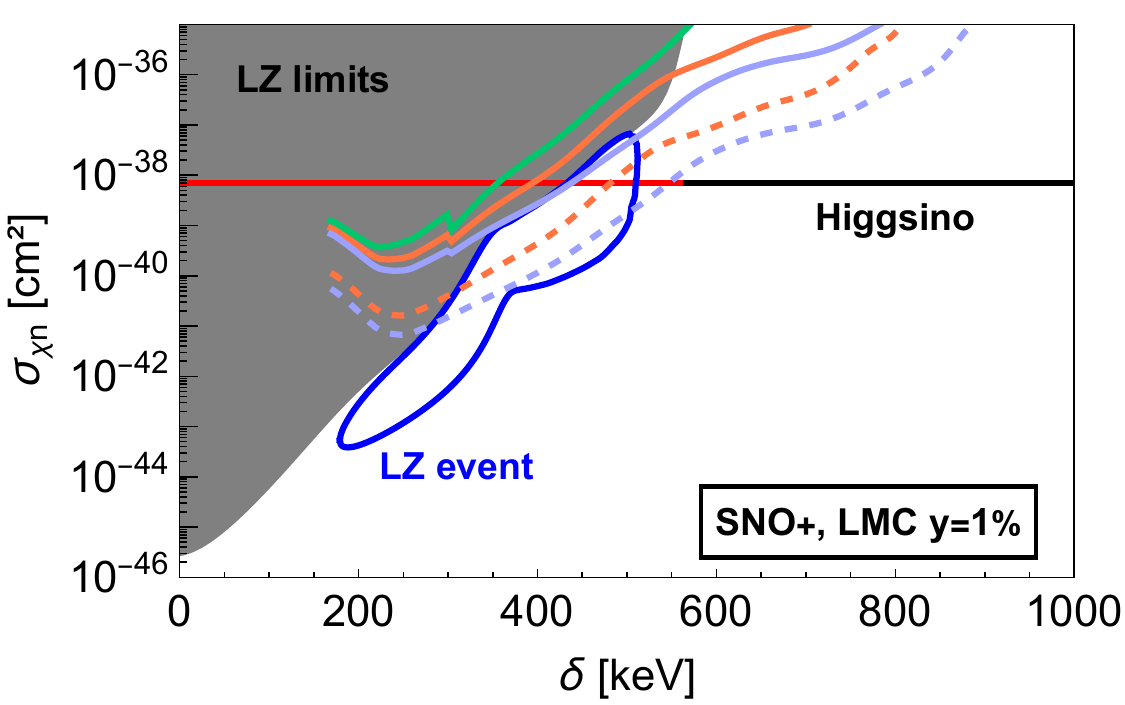}
        \caption{}
        \label{fig:sno_short100}
    \end{subfigure}
    
    \caption{Same as Fig.~\ref{fig:juno}, but for SNO+.}
    \label{fig:sno}
\end{figure*}

\begin{figure*}[t]
    \centering

    \begin{subfigure}{0.49\textwidth}
        \centering
        \includegraphics[width=\textwidth]{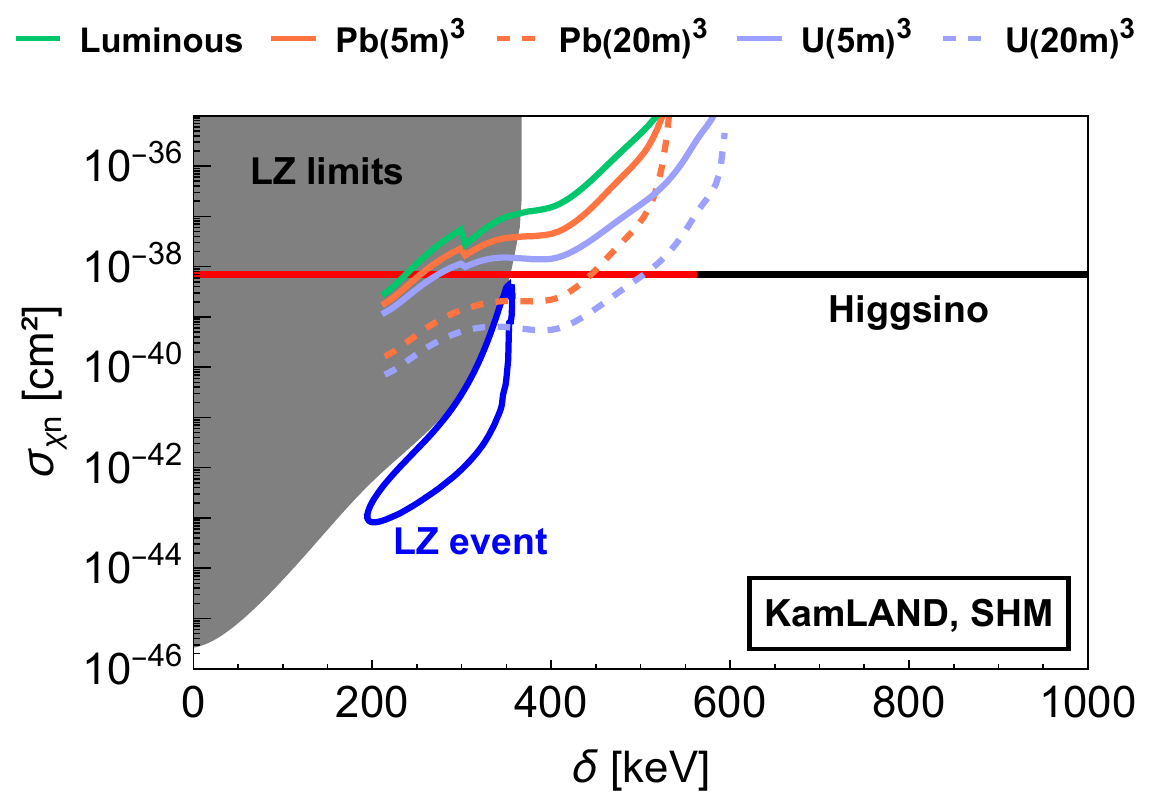}
        \caption{}
        \label{fig:kamlandSHM}
    \end{subfigure}
    \hfill
    \begin{subfigure}{0.49\textwidth}
        \centering
        \includegraphics[width=\textwidth]{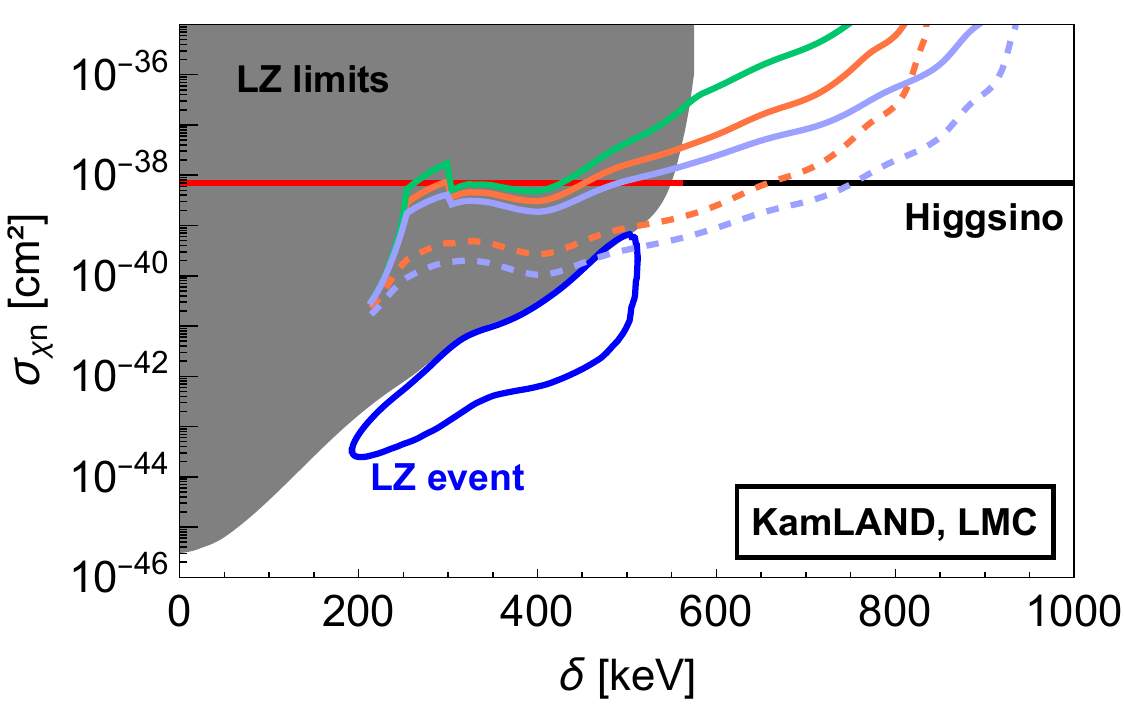}
        \caption{}
        \label{fig:kamlandLMC}
    \end{subfigure}
    
    \begin{subfigure}{0.49\textwidth}
        \centering
        \includegraphics[width=\textwidth]{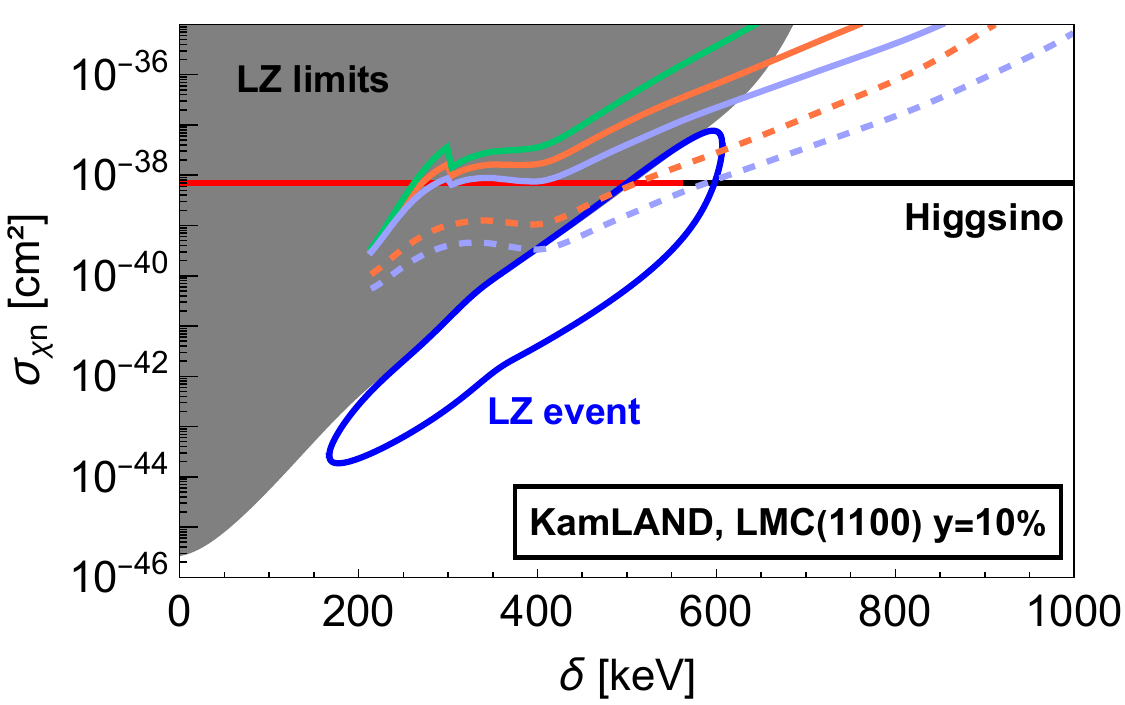}
        \caption{}
        \label{fig:kamland1100_short10}
    \end{subfigure}
    \hfill
    \begin{subfigure}{0.49\textwidth}
        \centering
        \includegraphics[width=\textwidth]{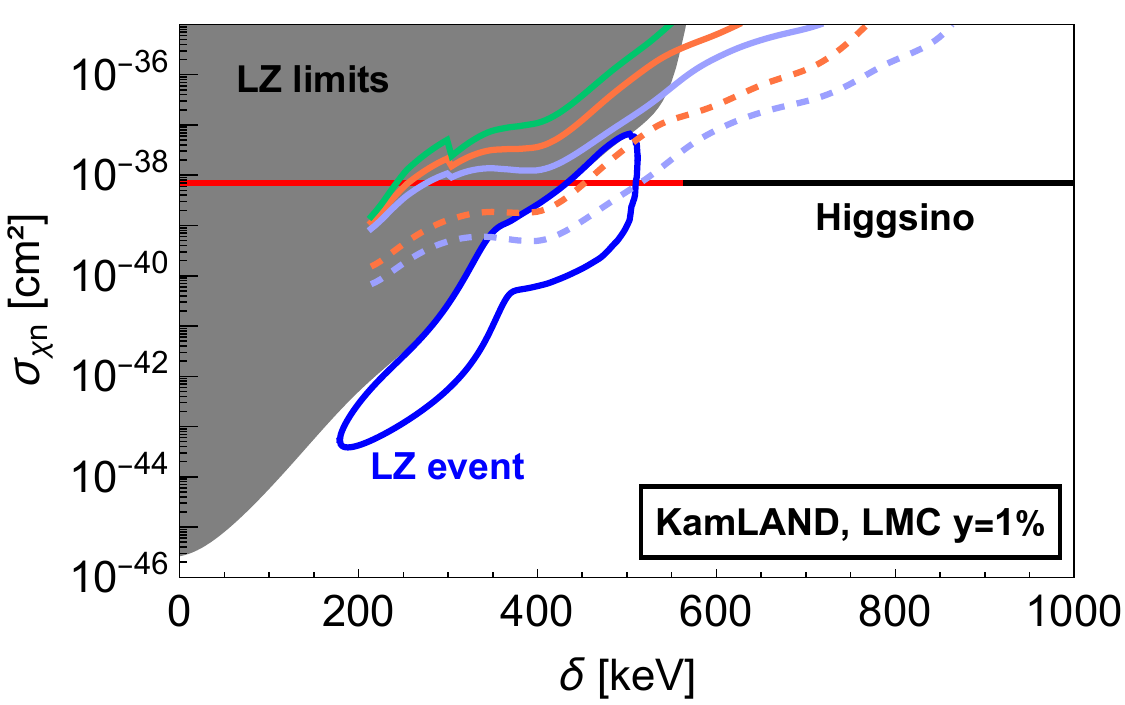}
        \caption{}
        \label{fig:kamland_short100}
    \end{subfigure}
    
    \caption{Same as Fig.~\ref{fig:juno}, but for KamLAND.}
    \label{fig:kamland}
\end{figure*}

\begin{figure*}[t!]
    \centering
    
    \begin{subfigure}{0.49\textwidth}
        \centering
        \includegraphics[width=\textwidth]{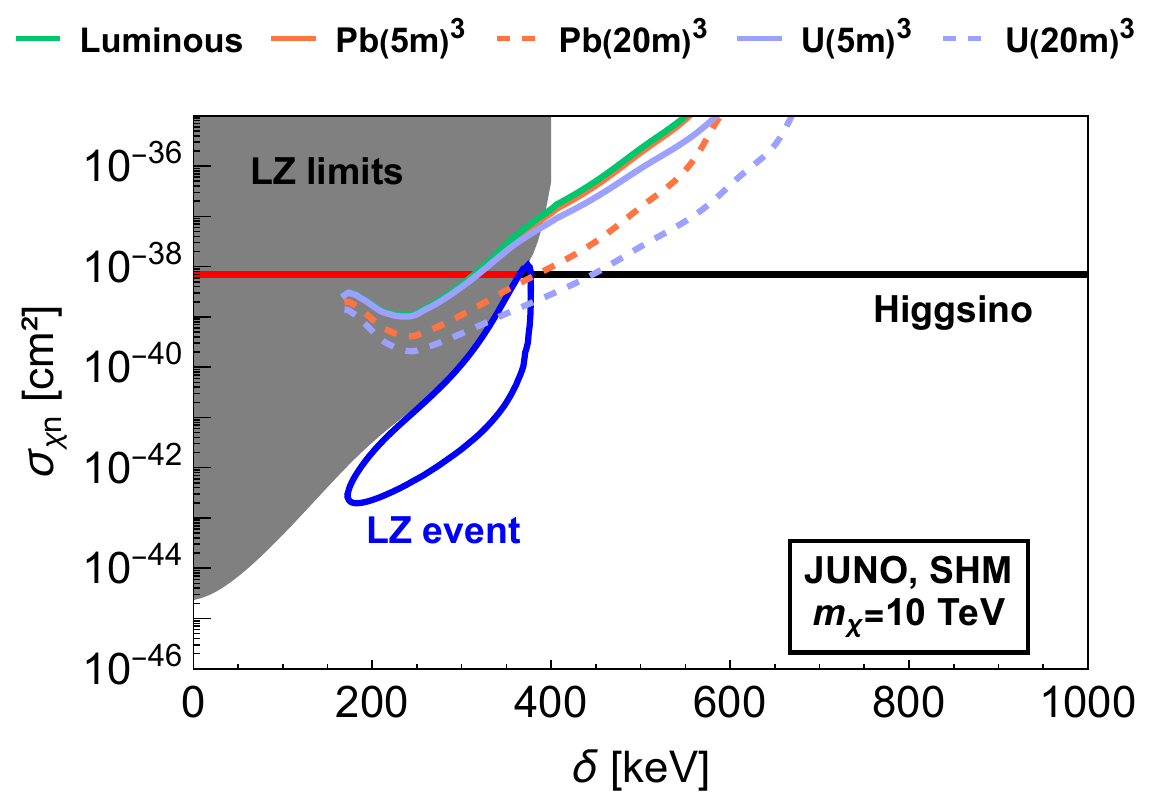}
        \caption{}
        \label{fig:junoSHM_10TeV}
    \end{subfigure}
    \hfill
    \begin{subfigure}{0.49\textwidth}
        \centering
        \includegraphics[width=\textwidth]{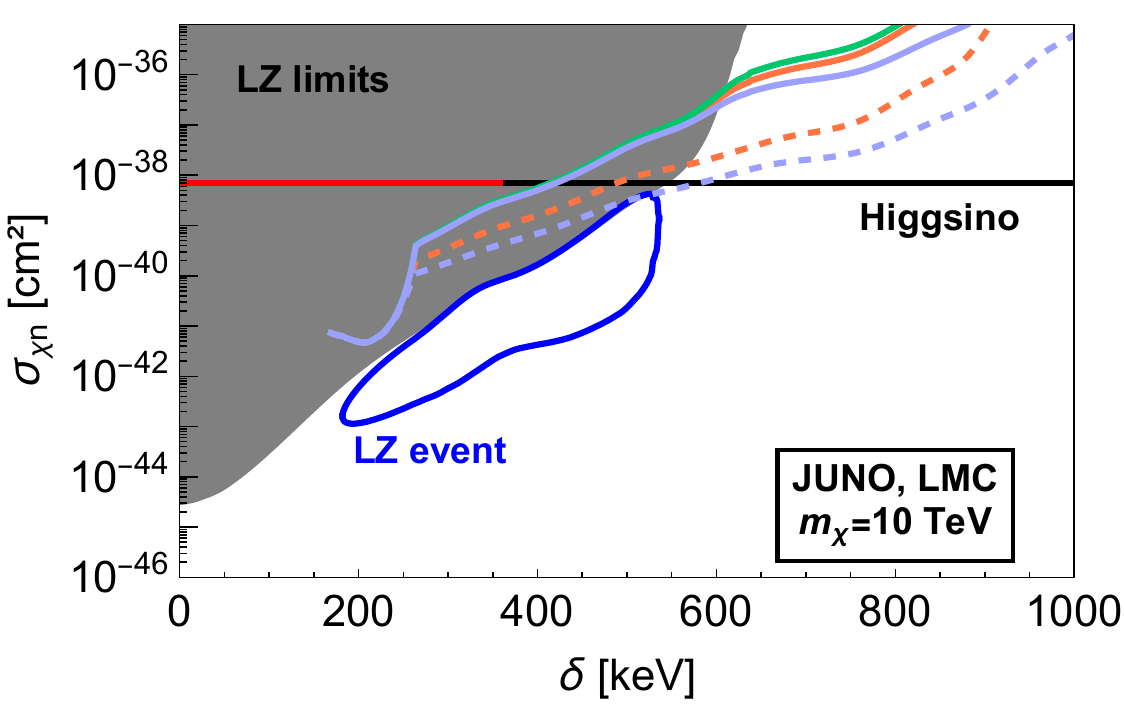}
        \caption{}
        \label{fig:junoLMC_10TeV}
    \end{subfigure}
    
    \begin{subfigure}{0.49\textwidth}
        \centering
        \includegraphics[width=\textwidth]{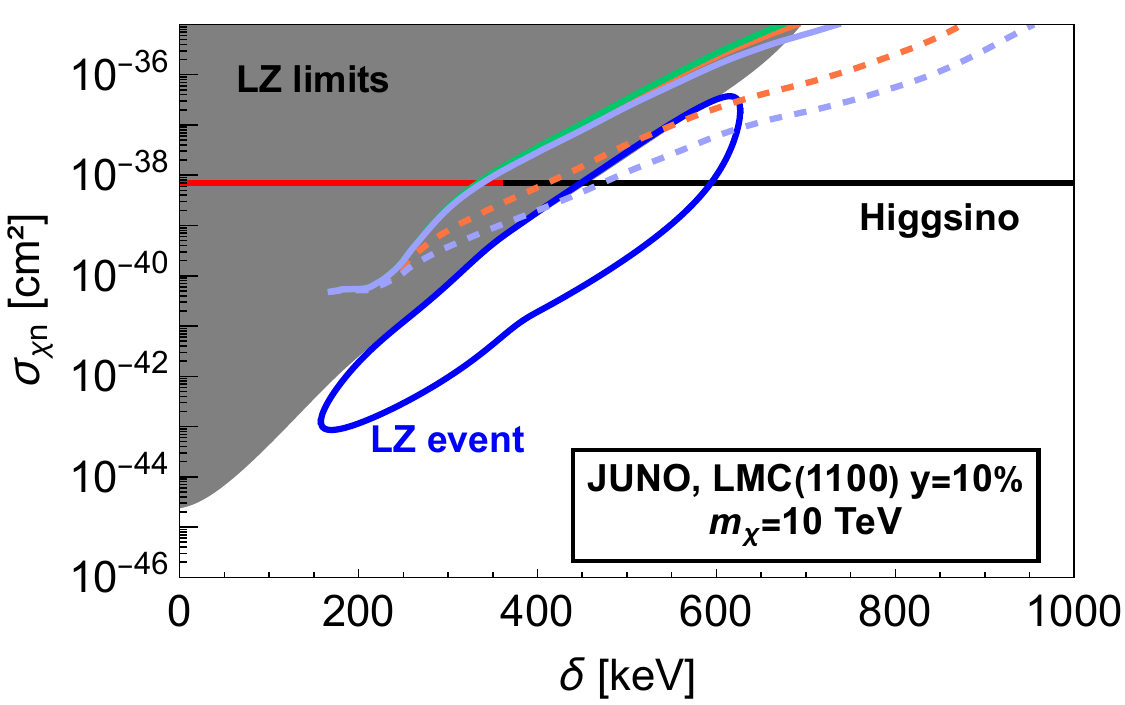}
        \caption{}
        \label{fig:juno1100_short10_10TeV}
    \end{subfigure}
    \hfill
    \begin{subfigure}{0.49\textwidth}
        \centering
        \includegraphics[width=\textwidth]{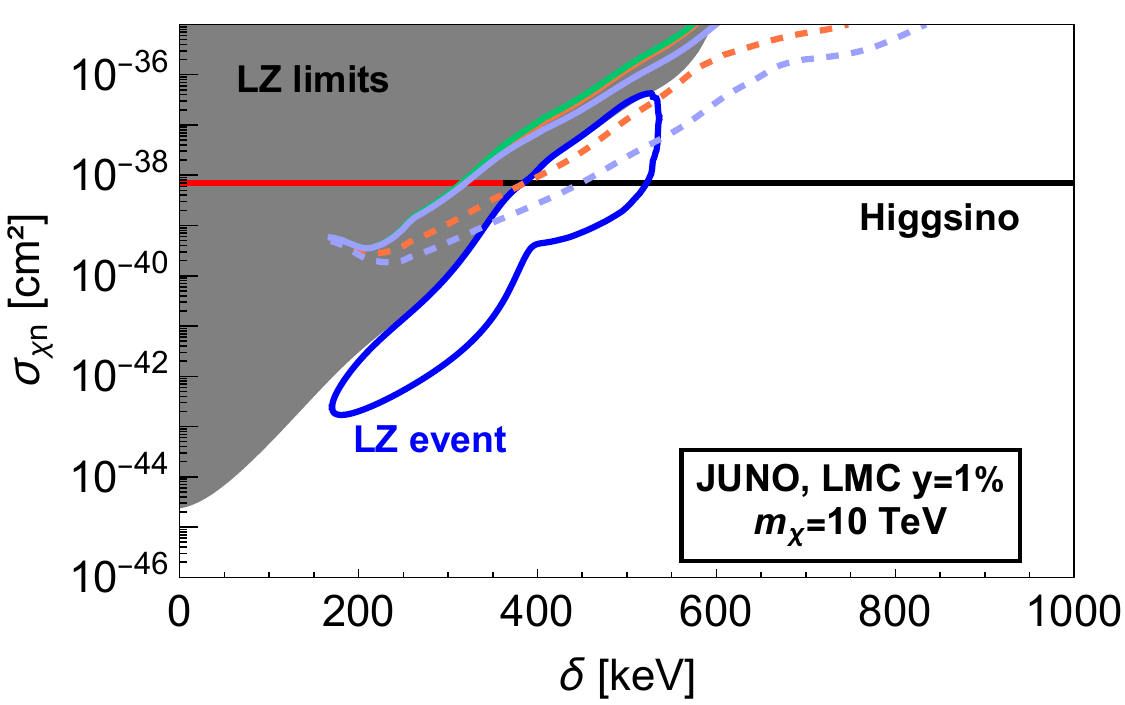}
        \caption{}
        \label{fig:juno_short100_10TeV}
    \end{subfigure}
    
    \caption{Same as Fig.~\ref{fig:juno}, but for $\mx=10$~TeV.}
    \label{fig:juno_10TeV}
\end{figure*}

We show the luminous detection results for SNO+~\cite{SNO+} ($R_D = 6$~m) in Fig.~\ref{fig:sno} and KamLAND~\cite{Kamland} ($R_D = 6.5$~m) in Fig.~\ref{fig:kamland}. We also show the result for JUNO for $\mx=10$~TeV in Fig.~\ref{fig:juno_10TeV}.

\clearpage
\bibliography{refs}

\end{document}